\documentclass{jfm}
\usepackage{bbm, dsfont} %police indicatrice
\usepackage{soul} % pour barrer
\usepackage{graphicx}
\usepackage{comment}

\usepackage{newtxtext}
\usepackage{newtxmath}
\usepackage{natbib}
\usepackage{abreviations}
\usepackage{hyperref}
\hypersetup{
 colorlinks = true,
 urlcolor = magenta,
 citecolor = magenta,
}

\newcommand{\RomanNumeralCaps}[1]
\linenumbers

\title{A {spatio-temporal} random synthetic turbulent velocity field: The underlying Gaussian structure}

\author{Matthieu Chatelain\aff{1}, J\'ulia Domingues Lemos\aff{1}, Wandrille Ruffenach\aff{1}, 
Micka\"el Bourgoin\aff{1}, Charles-Edouard Br\'ehier\aff{2}, Laurent Chevillard\aff{1,3} \corresp{\email{laurent.chevillard@ens-lyon.fr}}, Ilias Sibgatullin\aff{1}, Romain Volk\aff{1}}

\affiliation{\aff{1}Univ Lyon, Ens de Lyon, Univ Claude Bernard, CNRS, Laboratoire de Physique, 46 all\'ee d’Italie F-69342 Lyon, France\\
\aff{2} Universite de Pau et des Pays de l’Adour, E2S UPPA, CNRS, LMAP, Pau, France\\
\aff{3}CNRS, ICJ UMR5208, Ecole Centrale de Lyon, INSA Lyon, Universite Claude Bernard Lyon 1, Université Jean Monnet, 69622 Villeurbanne, France\\}

\begin{document}
\maketitle

\begin{abstract}

We develop, simulate and extend an initial proposition by \cite{ChaGaw03} concerning a random incompressible vector field able to reproduce key ingredients of three-dimensional turbulence in both space and time. In this article, we focus on the important underlying Gaussian framework. Presently, the statistical spatial structure of this velocity field is consistent with a divergence-free fractional Gaussian vector field that encodes all known properties of homogeneous and isotropic fluid turbulence at a given finite Reynolds number, up to second-order statistics. The temporal structure of the velocity field is introduced through a stochastic evolution of the respective Fourier modes. In the simplest picture, Fourier modes evolve according to an Ornstein-Uhlenbeck process, where the characteristic time scale depends on the wave-vector amplitude.  For consistency with direct numerical simulations (DNSs) of the Navier-Stokes equations, this time scale is inversely proportional to the wave vector amplitude. As a consequence, the characteristic velocity that governs the eddies is independent of their size and is related to the velocity standard deviation, which is consistent with some features of the so-called \textit{sweeping effect}. To ensure differentiability in time while respecting the {Markovian} nature of the evolution, we use the methodology developed by \cite{VigFri20} to propose a fully consistent stochastic picture. We finally derive analytically all statistical quantities in a continuous setup and develop precise and efficient numerical schemes of the corresponding periodic framework. Both exact predictions and numerical estimations of the model are compared to DNSs provided by the Johns Hopkins database. 
\end{abstract}


%{\bf MSC Codes } \textcolor{red}{{\it(Optional)} Please enter your MSC Codes here}

\section{Introduction}
\label{s.intro}

Fully developed fluid turbulence is emblematic of fluid mechanics and out-of-equilibrium statistical physics. It is usually observed in various natural flows and studied, in simpler situations than those of geophysical flows, in laboratory experiments and numerical simulations of the forced Navier-Stokes equations. An ancient tradition of measurements and observations has allowed the development of a very precise and self-consistent phenomenology capable of describing its statistical structure with few free parameters \citep{MonYag71,TenLum72,Fri95,Pop00}. From the theoretical side, the velocity of a turbulent fluid can be seen as the vanishing viscosity limit of the statistically stationary, isotropic, and homogeneous solution of the incompressible Navier-Stokes equations sustained by a random \citep{EswPop88}, possibly deterministic \citep{VinMen91}, force. Because of the interplay between non-linear and non-local aspects of this dynamics, with non-trivial influence of viscosity, the rigorous study of these solutions remains a fantastic mathematical challenge. The main purpose of this article is the presentation, extension, and simulation of a spatio-temporal turbulent velocity {which Fourier modes evolve according to a Markovian dynamics} in the simplest situation, {supplemented by additional underlying layers in a way that we will make precise}  in a more sophisticated version, able to reproduce several key statistical signatures. This stochastic model encompasses the spatial multi-scale nature of the observed turbulent velocity field, and a {associated}  random evolution in time.  In the infinite Reynolds number limit and when the analysis is restricted to second-order statistical laws, as captured in particular by a Gaussian framework, and assuming the model to evolve according to a {single-layer} Markovian dynamics, the present approach becomes equivalent to that proposed by \cite{ChaGaw03}. As far as we know, such models are unrelated to Navier-Stokes solutions, and we will therefore consider them as being synthetic, although they reflect important aspects of fluid turbulence.

Before presenting the proposition of \cite{ChaGaw03}, let us mention that synthetic turbulence has a long tradition in the literature, probably starting with fractional Brownian motions \citep{ManVan68}, known to have been initially proposed by \cite{Kol40} to describe the expected irregular nature of the velocity field, and later formulated in terms of spatial fields by \cite{Kra68}. Such fields, called fractional Gaussian fields, are at the core of turbulent phenomenology and give a precise random picture of the spatial structure of a turbulent velocity field, including in particular the famous power-law behavior of the power spectral density ``$k^{-5/3}$'', which is reminiscent of the asymptotic fractional regularity of the velocity field. Soon after proposing such random fields, \cite{Kra70} proposed a time-evolving version in which the time is included in the phase of the respective (spatial) Fourier modes. Such fields, referred to as \textit{Kinematic Simulations}, have then been popularized in a series of articles \citep{fung1992kinematic,fung1998two,malik1999lagrangian,CasPaz13} mostly devoted to the study of the dispersion of particle pairs. At the cost of introducing several free parameters that have yet to be physically interpreted, \cite{FavGod10} were able to reproduce in a realistic fashion several important aspects of the time evolution observed in direct numerical simulations (DNSs) of the Navier-Stokes equations while introducing a fluctuating nature in the dispersion relation relating wave numbers and frequencies. However, kinematic simulations have been acknowledged to be limited for certain aspects of particle dispersion \citep{ThoDev05,EyiBen13}, and, moreover, a precise dynamical formulation of this approach is still elusive. 

The purpose of the present article is to propose a random incompressible field that evolves according to a {stochastic evolution} consistent with the temporal structure of turbulence, which has been repeatedly observed, mostly in DNSs \citep{KanIsh99,CheRou05,FavGod10,CanRos17,GorBal21}. It is more difficult to observe this {evolution} in experimental measurements because it requires the absence of a mean flow, which is present in wind tunnels and jets. However, \cite{GorFal22} have conducted measurements of the temporal evolution of a flow stirred by magnetic particles able to generate turbulence without a mean flow, and temporal statistics were found to be similar to those observed in aforementioned numerical simulations. The temporal spectrum, defined as the variance of the temporal Fourier modes at a given frequency $\omega$, behaves in a universal fashion as a power-law ``$\omega^{-5/3}$'', with thus a similar exponent as observed on the spatial spectrum. Although understood and predicted by \cite{Ten75}, this power-law decay can be seen as counterintuitive and differs drastically from similarity arguments which would predict a much faster decrease, i.e. with a $-2$ power-law exponent, as has been observed in the Lagrangian framework, i.e. following fluid particles along their trajectory \citep{Yeu02,TosBod09,PinSaw12}. Temporal regularity of the velocity field is imposed by this power-law, which is thus the same as the spatial one, and turns out to have important consequences. In fact, the characteristic velocity of eddies of a given size is independent of that scale and turns out to be the one at large scale: this is the so-called \textit{sweeping effect}. 

From a statistical viewpoint, both this power-law decay of the temporal {power spectral density (PSD)} and the scale-independent characteristic velocity of eddies can be included in the model by setting a decorrelation time scale $T_\bk$ of Fourier modes proportional to the inverse wave-vector amplitude and whose multiplicative constant is inversely proportional to the velocity standard deviation allowing a consistent physical picture and proper dimension. This time scale dependent on $k$ is at the core of the random picture of \cite{ChaGaw03} {(see also earlier propositions of \cite{HolSig94}, \cite{KomPap97} and \cite{FanKom00}, and related discussions by  \cite{KomPes04} and \cite{EyiBen13}).} {Let us mention that, from a stochastic modeling point of view, building up a random field able to reproduce the observed aforementioned temporal structure is far from being obvious. In particular, it is known from earlier investigations by \cite{ManVan68} that fractional regularity usually requires a non Markovian dynamics and calls for sophisticated formulations of the fractional Brownian motions, as proposed by \cite{AloMaz00} and latterly by \cite{Che17}. To this regard, the proposition of  \cite{ChaGaw03} is very original, and shows that one could build instead the peculiar temporal regularity of the turbulent velocity in a Markovian way using the randomness introduced in the spatial dimensions using respective Fourier modes.  }

Nonetheless, it is well known that the turbulent velocity field is non-Gaussian, as can be seen by the behavior through scales of higher-order statistical quantities, such as third-order moment of velocity increments that quantifies the asymmetry of Probability density Functions (PDFs) and energy transfers, and the fourth-order one which focuses on the heavy-tail nature of these PDFs \citep{Fri95}. These non-Gaussian features of the velocity field are referred to as the \textit{intermittency phenomenon} which has a strong connection to multifractality. Several spatial random fields have been already proposed in the literature that are able to reproduce many aspects of this phenomenon, in particular the distributional nature of the dissipation field \citep{Yag62,Yag66,Man72,Man74,KahPey76,Kah85,MenSre87,MenSre91,RhoVar14}, and their extension to the velocity field \citep{JunLat94,BifBof98,ArnBac98a,BacDel01,RobVar08,CheRob10,BacDuv12,PerGar16,GraRou18,CheGar19,Muz19,CheLag20}, with generalizations towards structures and turbulent magnetic fields \citep{RosMen06,RosMen08,DurLes20,DurCha22,LubEff24} and solid-state physics \citep{LakPon25}. Let us mention that these aforementioned models are usually time-independent and they are meant to give a stochastic representation of statistically stationary solution of the Navier-Stokes equations. Recent works provide some generalization of these random fields with a given time evolution \citep{RenChe20,PeiTha23}. We keep the generalization of the present spatio-temporal random field to a multifractal framework for future works.

This article is organized as follows. In Chapter \ref{Sec:SpatoTempPhenoTurbulence}, we recall several important ingredients of the phenomenology of fluid turbulence based on second-order statistical quantities. Focusing on the spatial structure, we present in Section \ref{Sec:SpatialStats} basic statistical properties such as the {viscosity} independence of velocity variance and dissipation per unit of mass, the wave-vector dependence of the power spectral density and the respective scale dependence of the spatial structure functions. This allows us to define with precision several key parameters of the formulation, such as the mean H\"{o}lder exponent $H=1/3$, the Kolmogorov universal constant $C_2$ and the dissipative length scale $\eta_K$. These predictions are reformulated for a Gaussian model and compared with a DNS in Section \ref{Sec:CompSpatStatDNSModel}. We then discuss in Section \ref{Par:DiscussionSweeping} the behavior of characteristic time scales. In particular, we explore the implication of the sweeping effect, which has for statistical consequence that the characteristic turnover time scale of \textit{eddies} of a given size is independent of the scale and coincides with the large integral time scale. Following these observations, we deduce the implied statistical structure of the velocity field based on the second-order temporal structure function and the associated temporal PSD in Section \ref{Par:PresentModelOnR3}. We then present our spatio-temporal model in Chapter \ref{Sec:FormModel}. In Section \ref{Sec:FormModelIntroPerio}, we begin by reformulating former considerations with periodic boundary conditions. In order to achieve the statistical temporal structure for the description of turbulence, we propose in Section \ref{Sec:CausalEvolFM} a {Markovian}  stochastic evolution capable of generating such a field starting from any initial conditions. {In subsection \ref{Sec:DefFGF} we define} a Gaussian random field that mimics the spatial structure depicted in Sections \ref{Sec:SpatialStats} and \ref{Sec:FormModelIntroPerio}, which can be seen as a fractional Gaussian vector field. In subsection \ref{Sec:DefChavesGawOneLayer}, we recall the associated stochastic dynamics of \cite{ChaGaw03}, and generalize it to a smooth context, even toward an infinitely differentiable setup, in subsection \ref{Sec:GeneNLayers}. This will be achieved using the multi-layered evolution proposed by \cite{VigFri20}. An effective numerical schemes is proposed in Chapter \ref{Sec:NumericalSchemes}. We gather conclusions and perspectives in Chapter \ref{Sec:Conclusion}.  { Also, each step of the formulation will be compared to DNSs data, in both space and time. The numerical data provided by the Johns Hopkins Turbulence Database \citep{LiPer08} allow such a comparison. Preliminary analysis of this DNS database has been performed by \cite{Ren19}, and we will reproduce them here for convenience and completeness. We also provide key technical developments in the appendices.}

\section{The spatio-temporal statistical structure of a turbulent velocity field, and its continuous formulation on the whole space}\label{Sec:SpatoTempPhenoTurbulence}

\subsection{Basic ingredients of the phenomenology of fluid turbulence and the implied statistical spatial structure}\label{Sec:SpatialStats}

Let us consider the velocity vector field $\bu^\nu(t,\bx)\in\R^3$ for $\bx\in\R^3$ and $t\ge 0$, which reads component-wise $\bu^\nu=(u_i^\nu)_{1\le i \le 3}$, solution of the Navier-Stokes equations, for incompressible fluids of unit density,
\begin{align}\label{eq:NSforced}
\frac{\partial \bu^\nu}{\partial t}+(\bu^\nu\cdot \nabla) \bu^\nu = -\nabla p^\nu+\nu \Delta \bu^\nu +\mathbf{f},
\end{align}
where $\nu$ is the kinematic viscosity, $p^\nu (t,\bx)$ the pressure field obtained by the Poisson equation using the incompressibility constraint $\nabla\cdot \bu^\nu=0$, and $\mathbf{f}(t,\bx)$ a forcing vector assumed to be divergence-free, smooth in space, correlated on a characteristic length scale $L$ of the order of the integral length scale and independent of viscosity. We will denote the velocity gradient $\partial_j u_i^\nu$ of the component $u_i^\nu$ along $x_j$. 

For quite a general form of the forcing field $\bf{f}$, observations suggest that the velocity field $\bu^\nu$ reaches a statistically homogeneous and stationary regime, in which the velocity variance and average dissipation depend very weakly on viscosity, such that 
{\begin{align}\label{eq:AsymptVariance}
\sigma^2= \lim_{\nu\to 0} \E\left[|\bu^\nu|^2\right],
\end{align}}
and 
\begin{align}\label{eq:AsymptDiss}
\varepsilon=\lim_{\nu\to 0} \lim_{t\to \infty}\nu\sum_{ij}\E \left[\left( \partial_j u^\nu_i \right)^2\right],
\end{align}
are positive numbers. Starting from the randomly forced Navier-Stokes equations \eqref{eq:NSforced}, the expectations in \eqref{eq:AsymptVariance} and \eqref{eq:AsymptDiss} are taken over the  the forcing realisations. From a practical point of view, we will assume that expectations can be accurately replaced by empirical averages over space and/or time, which is in particular guaranteed by the statistical homogeneity and stationarity observed in DNSs and experiments. In the asymptotic limit of infinite Reynolds numbers, i.e. in the vanishing viscosity limit $\nu\to 0$, the finite quantities $\sigma$ \eqref{eq:AsymptVariance} and $\varepsilon$ \eqref{eq:AsymptDiss} are expected to be related according to the dimensional prediction {(see the classical textbooks by \cite{Fri95} and \cite{Pop00} for instance)}
\begin{align}\label{eq:AsymptVarDissRel}
\varepsilon \propto \frac{\sigma^3}{L}.
\end{align}

The independence of the velocity variance on viscosity \eqref{eq:AsymptVariance} is far from being obvious and is at the heart of the phenomenology of fluid turbulence. Recall that energy is injected into the fluid in a statistically stationary way through the forcing term $\bf{f}$ \eqref{eq:NSforced}. In order to maintain a velocity variance that becomes independent of viscosity \eqref{eq:AsymptVariance} as $\nu\to 0$, the fluid develops \textit{small scales} by populating Fourier modes located at higher wave numbers than those excited by the forcing. 

To characterize this behavior, usually referred to as the \textit{cascade} phenomenon, the {PSD} $E_\nu^{\text{\tiny{E}}}(\bk)$ has been measured and estimated in many studies. This quantity is defined to be the Fourier transform of the trace of the covariance tensor of the velocity vector field, that is - using Einstein’s summation convention over repeated indices -
\begin{align}\label{eq:DefPSD}
E_\nu^{\text{\tiny{E}}}(\bk) = \int_{\br\in\R^3} e^{-2i\pi \bk \cdot \br}\E \left[ u^\nu_i(t,\bx)u^\nu_i(t,\bx+\br) \right]\rd \br.
\end{align}
Note that the correlation $ \E \left[ u_i(t,\bx)u_i(t,\bx+\br) \right]$ is independent of both position $\bx$ and of time $t$ due to statistical homogeneity and stationarity respectively. Furthermore, by virtue of statistical isotropy, $E_\nu^{\text{\tiny{E}}}(\bk)=E_\nu^{\text{\tiny{E}}}(|\bk|)$, i.e. the PSD is a function of the norm $|\bk|$ only, and is independent of the direction of $\bk$. Note that the upper letter $^{\text{\tiny{E}}}$ indicates that $E_\nu^{\text{\tiny{E}}}(\bk)$ is a genuinely spatial, i.e. \textit{Eulerian}, quantity. As shown by \cite{Pop00}, the PSD $E_\nu^{\text{\tiny{E}}}(\bk)$ defined in \eqref{eq:DefPSD} fully determines the covariance function of the velocity field owing to the identity:
\begin{align}\label{eq:FullSpatCov}
 \E \left[ u^\nu_i(t,\bx)u^\nu_j(t,\bx+\br) \right] = \frac{1}{2}\int_{\bk\in\R^3} e^{2i\pi \bk \cdot \br} E_\nu^{\text{\tiny{E}}}(|\bk|) \widehat{P}_{ij}(\bk )\ \rd \bk ,
\end{align}
where Leray’s projector on divergence-free vector fields $ \widehat{P}_{ij}(\bk )$ has been used and is defined by
\begin{align}\label{eq:Leray}
 \widehat{P}_{ij}(\bk ) = \delta_{ij} - \frac{k_ik_j}{|\bk|^2},
\end{align}
with $\delta_{ij} $ the Kronecker delta symbol.

Although being accessible in Direct Numerical Simulations (DNSs) of the Navier-Stokes equations, the 3$d$ power spectral density $E_\nu^{\text{\tiny{E}}}(\bk)$ \eqref{eq:DefPSD} is difficult to obtain from experimental measurements. Instead, one-dimensional \textit{longitudinal} power spectral density has traditionally been estimated. By virtue of statistical isotropy, this quantity is defined as the one-dimensional Fourier transform of the correlation function of one of the component of velocity, say $u_1$, along the direction $\hat{e}_1$, that is: \begin{align}\label{eq:DefPSD1dLong}
E_\nu^{\text{\tiny{E,long}}}(k) = \int_{\ell\in\R} e^{-2i\pi k\ell}\E \left[ u^\nu_1(t,\bx)u^\nu_1(t,\bx+\ell \hat{e}_1) \right]\rd\ell.
\end{align}
It can be checked that $E_\nu^{\text{\tiny{E,long}}}(k)$ is a positive and even function of $k$. The three-dimensional PSD $E_\nu^{\text{\tiny{E}}}(\bk)$ \eqref{eq:DefPSD}, which is a function of the norm $k = |\bk|$ of the wave vector $\bk$, and the one-dimensional longitudinal PSD $E_\nu^{\text{\tiny{E,long}}}(k)$ \eqref{eq:DefPSD1dLong} are related as follows: 
\begin{align}\label{eq:RelPSD1dLong3d}
E_\nu^{\text{\tiny{E}}}(k) = \frac{k}{2\pi}\frac{\rd}{\rd k}\left(\frac{1}{k}\frac{\rd E_\nu^{\text{\tiny{E,long}}}(k)}{\rd k }\right).
\end{align}
The above identity is a consequence of incompressibility and statistical isotropy. A complete proof of the relation \eqref{eq:RelPSD1dLong3d} is provided by \cite{Pop00}.

Going back to the aforementioned phenomenological aspects of fluid turbulence, a manifestation of the underlying cascading process of energy can be read on the PSD, which will follow, in the limit of infinite Reynolds numbers, a power-law behavior at large wave numbers. This is known as the Kolmogorov ``$k^{-5/3}$'' spectrum \citep{Fri95}. At a given finite Reynolds number, a realistic parameterization of the longitudinal PSD $E_\nu^{\text{\tiny{E,long}}}(k)$ \eqref{eq:DefPSD1dLong} would require a regularization at small wave numbers of order $1/L$, i.e. around the inverse integral length scale stemming from the forcing term, and a dissipative cut-off at large wave numbers $1/\eta_K$, where {$\eta_K \propto (\nu^3/\varepsilon)^{1/4}$}is known as Kolmogorov's dissipative length scale. Sophisticated models that do so have been proposed in the literature \citep{Pop00,MeyMen08}. For the sake of simplicity, we will be working with the following simple but realistic parameterization:
\begin{align}\label{eq:ModelPSD1dLong}
E_\nu^{\text{\tiny{E,long}}}(k) =D_2 |k|_{1/L}^{-5/3} e^{-\eta_d k},
\end{align}
where we have introduced a regularized norm $|k|_{1/L} = \sqrt{k^2 + 1/L^2}$, a multiplicative constant $D_2>0$ which is related to velocity variance \eqref{eq:AsymptVariance} according to the identity
\begin{align}\label{eq:RelD2VelVariance}
D_2 = \frac{\sigma^2}{3L^{2/3}}\frac{\Gamma(5/6)}{\sqrt{\pi}\Gamma(1/3)},
\end{align}
and
\begin{align}\label{eq:DefEtaD}
\eta_d = 2\pi \left( \frac{10C_2\Gamma(4/3)}{\Gamma(1/3)}\right)^{3/4}\left(\frac{\nu^3}{\varepsilon}\right)^{1/4},
\end{align}
which is of order of the Kolmogorov dissipative length scale $\eta_K$. $\Gamma\lp z\rp$ denotes here the gamma function. The expression of the pre-factor in the definition of $\eta_d$ \eqref{eq:DefEtaD} is derived below{, and is chosen such that to obtain ultimately \eqref{eq:DefS2Holder13LongWithCk}.} The so-called Kolmogorov constant $C_2$ has been estimated on experimental data and is close to $C_2 \approx 2$, see \cite{Pop00}. From the proposed expression for $E_\nu^{\text{\tiny{E,long}}}(k)$ \eqref{eq:ModelPSD1dLong}, we deduce from the identity \eqref{eq:RelPSD1dLong3d} the expression of the three-dimensional PSD $E_\nu^{\text{\tiny{E}}}(k)$ \eqref{eq:DefPSD}, which has a particularly simple expression in the vanishing viscosity limit, i.e.
\begin{align}\label{eq:DeducedModelPSD3d}
E^{\text{\tiny{E}}}(k) =\lim_{\nu\to 0}E_\nu^{\text{\tiny{E}}}(k) = \frac{1}{2\pi}\frac{55}{9} D_2 k^2 |\bk|_{1/L}^{-17/3}.
\end{align}

Note that the proportionality constant appearing in the expression of $D_2$ \eqref{eq:RelD2VelVariance} ensures that the velocity variance $\sigma^2$ \eqref{eq:AsymptVariance} is given by the integral of the PSD \eqref{eq:ModelPSD1dLong}.  We therefore have the identity
\begin{align}\label{eq:VarianceFromPSD}
\sigma_\nu^2=\lim_{t\to \infty}\E\left[|\bu^\nu(t,\bx)|^2\right]=\int_{\bk\in\R^3} E_\nu^{\text{\tiny{E}}}(\bk) \rd \bk=3\int_{k\in\R} E_\nu^{\text{\tiny{E,long}}}(k) \ \rd k .
\end{align}
Consequently, using the limiting value of $\sigma_\nu^2$ as $\nu\to 0$ as expressed in \eqref{eq:VarianceFromPSD} and the parameterization of the longitudinal PSD provided in \eqref{eq:ModelPSD1dLong} and recalling that $\eta_d$ vanishes as $\nu\to 0$ \eqref{eq:DefEtaD}, one obtains:
{\begin{align}\label{eq:VarianceAndD2}
\sigma^2=\lim_{\nu\to 0}\sigma_\nu^2=6D_2\int_{k=0}^{\infty} |k|_{1/L}^{-5/3} \ \rd k ,
\end{align}}
where $D_2$ is given by \eqref{eq:RelD2VelVariance}. Also, as claimed above, the regularization parameter $L$ appearing in the expression of the longitudinal PSD \eqref{eq:ModelPSD1dLong} can be interpreted as the correlation length scale of the longitudinal component. More precisely, $L$ can be related to the so-called integral length scale $L_{\text{\tiny{int}}}$, i.e. the integral of the correlation function, as follows:
\begin{align}\label{eq:RelatLtoLint}
L_{\text{\tiny{int}}}\equiv\int_{\ell>0}\frac{3}{\sigma^2}\E \left[ u^\nu_1(t,\bx)u^\nu_1(t,\bx+\ell \hat{e}_1) \right]d\ell = \frac{3}{2\sigma^2}E_\nu^{\text{\tiny{E,long}}}(0) = \frac{\Gamma(5/6)}{2\sqrt{\pi}\Gamma(1/3)}L.
\end{align}

Concerning the average dissipation $\varepsilon_\nu$, by making use of the isotropy condition, we have
\begin{align}\label{eq:AverageDissFromPSD}
\varepsilon_\nu=15\nu\E \left[\left( \partial_1 u_1 \right)^2\right]&=15\nu\int_{k\in\R} (2\pi)^2k^2E_\nu^{\text{\tiny{E,long}}}(k) \rd k\notag\\
&=30(2\pi)^2D_2\nu\eta_d^{-4/3}\int_{k=0}^{\infty} k^2|k|_{\eta_d/L}^{-5/3} e^{-k}\rd k\notag\\
&\build{\sim}_{\nu\to 0}^{}30(2\pi)^2D_2\nu\eta_d^{-4/3}\Gamma(4/3).
\end{align}
Taking into account the expressions of $D_2$ \eqref{eq:RelD2VelVariance} and $\eta_d$ \eqref{eq:DefEtaD}, we then obtain
\begin{align}\label{eq:AverageDissFromPSDLimiting}
\varepsilon=\lim_{\nu\to 0}\varepsilon_\nu = 2\pi \left( \frac{\Gamma(5/6)}{\sqrt{\pi}C_2}\right)^{3/2}\frac{\sigma^3}{L}.
\end{align}

In the  vanishing viscosity limit, a Gaussian random field which possesses a PSD that decays as a power-law of the type proposed in \eqref{eq:ModelPSD1dLong} is known in the literature as a fractional Gaussian field of H\"{o}lder (or Hurst) parameter $H=1/3$. In particular, such a field is continuous and its variance is finite \eqref{eq:VarianceAndD2}, but it is nowhere differentiable, and must be regarded as being H\"{o}lder-continuous. To characterize the H\"{o}lder-continuity, instead of using gradients that are expected to be infinite, one needs to consider velocity increments that are well-posed. Precisely, the longitudinal velocity increment $\delta_\bll^{\text{\tiny{long}}} \bu^\nu$ is defined by
\begin{align}\label{eq:DefVelIncrVectLong}
\delta_\ell^{\text{\tiny{long}}} \bu^\nu(t,\bx) = \left(\bu^\nu(t,\bx+\bell) - \bu^\nu(t,\bx)\right)\cdot \frac{\bell}{\ell} \ ,
\end{align}
for all $\bll \neq 0$. The variance of the longitudinal increment \eqref{eq:DefVelIncrVectLong} is known as the longitudinal second-order structure function, and is given by
\begin{align}\label{eq:DefS2FiniteVisc}
S_{2,\nu}^{\text{\tiny{long}}}(\ell)=\E \left[ \left(\delta_\ell^{\text{\tiny{long}}} \bu^\nu\right)^2\right].
\end{align}
In the vanishing viscosity limit, one obtains
\begin{align}\label{eq:DefS2Holder13Long}
S_2^{\text{\tiny{long}}}(\ell) &= \lim_{\nu\to 0} \E \left[ \left(\delta_\ell^{\text{\tiny{long}}} \bu^\nu\right)^2\right] \notag \\
&= 2\int_{k\in\R} \left( 1-e^{2i\pi k\ell}\right) E^{\text{\tiny{E,long}}}(k) \rd k = 4D_2\int_{k=0}^{\infty} \left( 1-\cos(2\pi k\ell)\right) |k|_{1/L}^{-5/3} \rd k\notag\\
&\build{\sim}_{\ell\to 0}^{} 4D_2 \ell^{2/3}\int_{k=0}^{\infty} \left( 1-\cos(2\pi k)\right) k^{-5/3} \rd k=3D_2(2\pi)^{2/3}\Gamma(1/3)\ell^{2/3}.
\end{align}

As can be seen from the behavior of $S_2^{\text{\tiny{long}}}(\ell)$ \eqref{eq:DefS2Holder13Long} on small scales $\ell\to 0$, the second-order structure function exhibits a power-law behavior with respect to the scale $\ell$ with the exponent $2/3$. Using then the expression of $D_2$ \eqref{eq:RelD2VelVariance}, and expressing the remaining multiplicative constant in units of the average dissipation $\varepsilon$ \eqref{eq:AverageDissFromPSDLimiting}, we finally get
\begin{align}\label{eq:DefS2Holder13LongWithCk}
S_2^{\text{\tiny{long}}}(\ell) \build{\sim}_{\ell \to 0}^{} C_2|\varepsilon \ell|^{2/3},
\end{align}
that justifies the introduction of the Kolmogorov constant $C_2$ in the definition of the dissipative cutoff $\eta_d$ \eqref{eq:DefEtaD}.

More generally, let us go back  to the full vector field. Its covariance is provided in \eqref{eq:FullSpatCov}. Let us introduce the increment of a given velocity component $u_i^\nu$
\begin{align}\label{eq:DefVelocityIncrementCompI}
\delta_{\ell}u_i^\nu (t,\bx) = u_i^\nu (t,\bx+\bell)-u_i^\nu (t,\bx),
\end{align}
and the respective second-order structure function
\begin{align}\label{eq:DefSijNuTo0}
S_{2,ij}(\ell) = \lim_{\nu\to 0}\E\left[\delta_{\ell}u_i^\nu (t,\bx) \delta_{\ell}u_j^\nu (t,\bx) \right].
\end{align}
Then,  under the assumption of statistical isotropy and incompressibility, from \citep{Pop00}, one obtains 
\begin{align}\label{eq:SijIsotropyIncomp}
S_{2,ij}(\ell) \build{\sim}_{\ell \to 0}^{} C_2|\varepsilon \ell|^{2/3}\left[-\frac{1}{3}\frac{\ell_i\ell_j}{\ell ^2}+\frac{4}{3}\delta_{ij} \right].
\end{align}
As a consequence, one has  
\begin{align}\label{eq:DefS2Holder13}
S_2(\ell)\equiv S_{2,ii}(\ell) = \lim_{\nu\to 0} \E \left[ \left|\delta_\ell \bu^\nu\right|^2\right] &= 2\int_{\bk\in\R^3} \left( 1-e^{2i\pi \bk \cdot \bell}\right) E^{\text{\tiny{E}}}(|\bk|) \rd \bk\notag\\
&\build{\sim}_{\ell \to 0}^{}\frac{11}{3}C_2|\varepsilon \ell|^{2/3}.
\end{align}
Firstly taking the limit $\ell\to 0$, and secondly the limit $\eta_d\to 0$ (or equivalently $\nu\to 0$), we obtain then the power-law behavior
\begin{align}\label{eq:TaylorS2Long}
\E \left[ \left(\delta_\ell^{\text{\tiny{long}}} \bu^\nu\right)^2\right] \build{\sim}_{\ell \to 0}^{} 2\ell ^2\int_{k\in\R} \left(2\pi k \right)^2 E^{\text{\tiny{E,long}}}_\nu(k) \rd k\build{\sim}_{\nu\to 0}^{} \frac{\varepsilon_\nu}{15\nu} \ell ^2,% \notag
\end{align}
where the expression of $\varepsilon_\nu$ is given in \eqref{eq:AverageDissFromPSD}. The power-law \eqref{eq:TaylorS2Long} is associated with differentiable velocity fields instead of being H\"{o}lder continuous with fractional exponent $H=1/3$.

\subsection{A first comparison concerning the spatial structure of DNS and the Gaussian model}\label{Sec:CompSpatStatDNSModel}

\begin{figure}
\centering
\includegraphics[scale = 0.7]{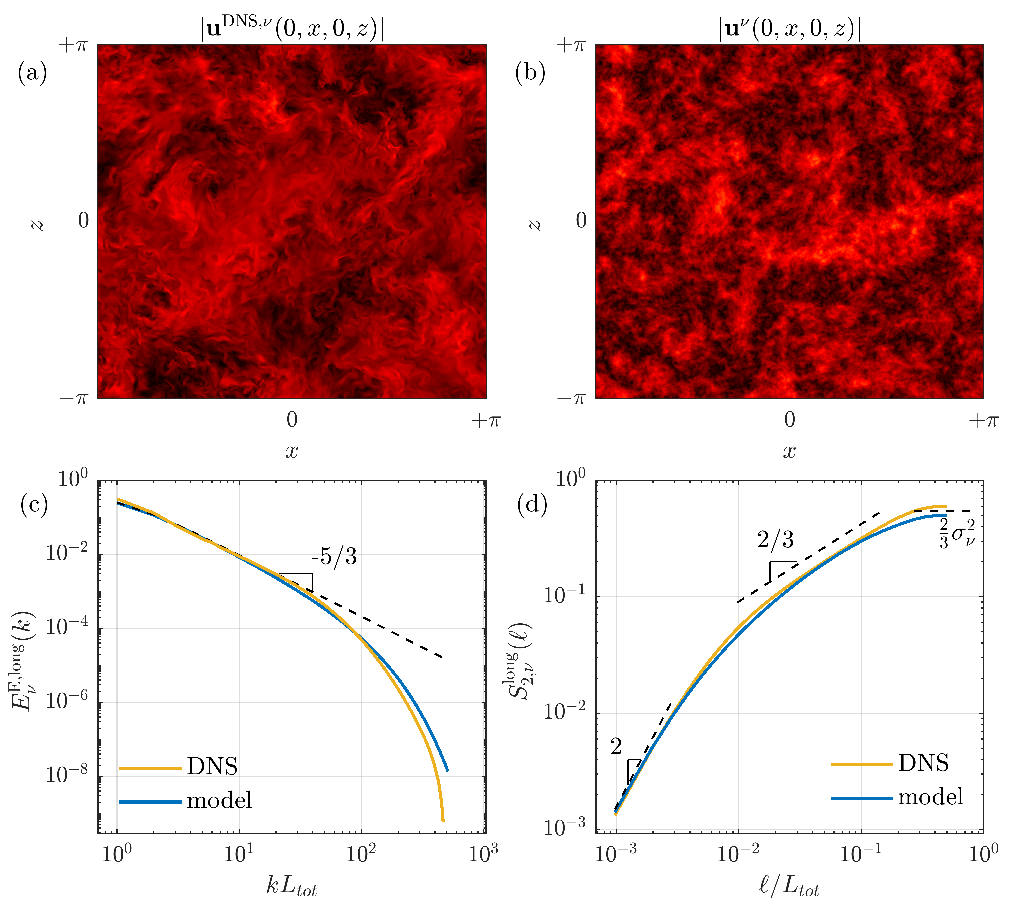}
\caption{Comparison of the instantaneous and statistical spatial structure of a DNS velocity field $\bu^{\text{\tiny{DNS}},\nu}(t,\bx)$ and of a realization of the model which coincides with a fractional Gaussian vector field at a given instant of time in the statistically stationary regime. (a) Instantaneous spatial profile of the norm of the DNS velocity field $|\bu^{\text{\tiny{DNS}},\nu}(t,\bx)|$ in the plane $y=0$, at the initial time of the DNS dataset. (b) Same as (a) for the model, we used the same colorbar in both representations. (c) Estimation of the longitudinal PSDs $E_\nu^{\text{\tiny{E,long}}}(k)$ \eqref{eq:DefPSD1dLong} based on the variance of the Fourier modes of the one-dimensional discrete Fourier transform. Statistics (DNS in orange and model in blue) are estimated by averaging both in space and time. We superimpose with a black dashed line the inviscid limit of the functional form provided in \eqref{eq:ModelPSD1dLong}, corresponding to the power-law $D_2|\bk|^{-5/3}$. (d) Statistical estimation of the second-order longitudinal structure function 
$S_{2,\nu}^{\text{\tiny{long}}}(\ell)$ \eqref{eq:DefS2FiniteVisc}, same colors as (c). We superimpose with dashed lines the inviscid predictions in the three ranges of scales of interest: (i) at large scales of order $L_{\text{\tiny{tot}}}$, $S_2^{\text{\tiny{long}}}(\ell)$ reaches the plateau $\frac{2}{3}\sigma^2$, where $\sigma^2$ is related to the free parameter $D_2$ according to \eqref{eq:RelD2VelVariance}, (ii) in the intermediate inertial range of scales, we represent the prediction made in \eqref{eq:DefS2Holder13LongWithCk}, and (iii) by the smooth behavior $\propto \ell^2$ in the dissipative range, as it is predicted in \eqref{eq:TaylorS2Long}. }
\label{Fig:SpatialSnaps}
\end{figure}

We show in Fig. \ref{Fig:SpatialSnaps} a comparison between the velocity field $\bu^{\text{\tiny{DNS}},\nu}(t,\bx)$ (Fig. \ref{Fig:SpatialSnaps}(a)) extracted from a direct numerical simulations (DNSs) of the forced Navier-Stokes equations \eqref{eq:NSforced} and a Gaussian model (Fig. \ref{Fig:SpatialSnaps}(b)). Concerning the DNS, it is provided by the Johns Hopkins turbulence database \citep{LiPer08}. The data set concerns a simulation of the Navier-Stokes equations in a homogeneous and isotropic situation, using a pseudo-spectral method fully de-aliased with $N=1024$ nodes in each direction, over 5028 timesteps, corresponding to roughly speaking ten turnover time scales. The Taylor-based Reynolds number is estimated to be 433, corresponding to a fully developed turbulent state. Detailed properties of the simulation are provided on their website. For our purpose, we have downloaded the three velocity components over three spatial slices : $(0,y,z)$, $(x,0,z)$ and $(x,y,0)$, at all available times. The Gaussian model $\bu^{\text{\tiny{fGv}},\nu}(\bx)$ corresponds to a fractional Gaussian vector field, in a periodic setting, presented in subsection \ref{Sec:DefFGF}. At this stage, it is enough to state that this is the unique statistically homogeneous, isotropic and incompressible Gaussian vector field consistent with the longitudinal PSD 
$E_\nu^{\text{\tiny{E,long}}}(k)$ provided in \eqref{eq:ModelPSD1dLong}. Following a suitable truncation of the Fourier modes, as described in Section \ref{Sec:DFT}, we have similarly performed a simulation of this Gaussian vector field on a periodic box of size $L_{\text{\tiny{tot}}}=2\pi$  using $N=1024$ nodes in each direction, with the remaining free parameters entering in \eqref{eq:ModelPSD1dLong} chosen to match the DNS dataset. This corresponds to $D_2 = 0.021$, $L= L_{\text{\tiny{tot}}} $ and $\eta_d = 0.085$. The results of our comparison are collected in Fig. \ref{Fig:SpatialSnaps}. 

We begin by displaying in Fig. \ref{Fig:SpatialSnaps}(a) a snapshot of the norm of the DNS velocity field $|\bu^{\text{\tiny{DNS}},\nu}(t = 0,x,y=0,z)|$ in the plane defined by the Cartesian coordinates $y=0$ at the initial time of the dataset, which has been chosen in the statistically stationary regime. Similarly, in Fig. \ref{Fig:SpatialSnaps}(b) we display for comparison the same snapshot but for the model $|\bu^\nu(t=0,x,y=0,z)|$. The colorbars are the same in both panels, dark corresponding to the lowest values and light corresponding to the highest values of the norm. We observe similar trends concerning amplitude of the fluctuations, and also the statistical symmetries, i.e. homogeneity and isotropy. However, it appears that, while $\bu^{\text{\tiny{DNS}},\nu}(t,\bx)$ exhibits filamentary structures, the Gaussian field looks more patchy. Despite their different appearances, we will show that these two vector fields have consistent second-order statistics.

The statistical analysis of these two vector fields is shown in panels \ref{Fig:SpatialSnaps}(c) and \ref{Fig:SpatialSnaps}(d). The respective longitudinal PSDs $E_\nu^{\text{\tiny{E,long}}}(k)$ are shown in (c). Concerning the DNS, it is obtained as the variance of the respective Fourier modes obtained by performing one-dimensional discrete Fourier transforms of the longitudinal components, say, for instance, the $x$ component of velocity along the direction $x$. The statistical sample is averaged over the three longitudinal components, spatially over the slices, and over time. Results of our estimation are displayed with an orange solid line (light gray). The estimation of the Gaussian model statistics is also performed by averaging over the three components, over space and time, as has been done for the DNS. We display the result of our estimation using a solid blue line (dark gray). As we can observe, both longitudinal PSDs superimpose in a very satisfactory manner, showing that, based on the variance of Fourier modes, these two fields are statistically {almost} indiscernible. We mention that we could have, instead of the estimate on the Gaussian model, displayed the theoretical functional form provided in \eqref{eq:ModelPSD1dLong} without noticeable difference, as expected. The result of our statistical estimation of the longitudinal structure function $S_{2,\nu}^{\text{\tiny{long}}}(\ell)$ \eqref{eq:DefS2FiniteVisc} is displayed in Fig. \ref{Fig:SpatialSnaps}(d), using an orange line (light gray) for DNS and a blue line (dark gray) for the Gaussian model, averages have been taken on the same statistical sample as the one considered for Fig. \ref{Fig:SpatialSnaps}(c). The three ranges of scales, including decorrelation length scales of order of the size of the box $L_{\text{\tiny{tot}}}$, inertial and dissipative length scales, are reproduced in a very satisfactory manner by the model, as expected from the behaviors of the PSDs of Fig. \ref{Fig:SpatialSnaps}(c).

We conclude this section noting that, whereas it is clear that instantaneous snapshots of DNS and spatial representations of the model, as they are displayed in Figs. \ref{Fig:SpatialSnaps}(a) and \ref{Fig:SpatialSnaps}(b), exhibit differences regarding in particular filamentary structures, they are nonetheless barely indiscernible from a second-order statistical point of view. For this reason, we claim that a model based on a Gaussian approximation is a fairly good starting point. We are planning to improve this in future investigations.

\subsection{The statistical temporal structure of the turbulent velocity field and the sweeping effect}\label{Par:DiscussionSweeping}

Fluctuations in time of turbulent velocity can be considered counter-intuitive from a dimensional point of view \citep{TenLum72,Ten75}. In particular, the sweeping effect that can be seen, loosely speaking, as the advection of small scales by large scales, has been investigated and defined in very different ways by several groups \citep{Krai64,BelLvo87,GotRog93,KanIsh99,ChaGaw03,CheRou05,FavGod10,BifCal11,DriJoh17,CanRos17,GorBal21,MatOts21}. We should bear in mind that such a mechanism should not be confused with Taylor's hypothesis related to the presence of a mean flow \citep{WilNar12,HeJi17}. As far as we are concerned, let us discuss this phenomenon in the spirit of \cite{Ten75} which focuses on the dimensional aspects of this phenomenon and its statistical signature.

From the statistical spatial structure of the velocity field formerly presented, in large part given by the scaling relation provided in \eqref{eq:DefS2Holder13LongWithCk} and \eqref{eq:DefS2Holder13}, let us infer the statistical temporal structure of the velocity field $\bu^\nu$ in the asymptotic limit $\nu\to 0$. The H\"{o}lder regularity pinpointed in \eqref{eq:DefS2Holder13LongWithCk} and \eqref{eq:DefS2Holder13} suggests that eddies of characteristic length scale $\ell $ have a characteristic turnover time scale $\tau_{\text{e}}(\ell )$ given by
\begin{align}\label{eq:DefTauEddies}
\tau_{\text{e}}(\ell ) \equiv \frac{\ell }{\sqrt{S_2(\ell)}}\build{\propto}_{\ell \to 0}^{} \frac{\ell ^{2/3}}{\sqrt{C_2}\varepsilon^{1/3}}.
\end{align}
The square-root of the structure function $\sqrt{S_2(\ell)}$ can be interpreted as the typical velocity of eddies of size $\ell $. The characteristic turnover time $\tau_{\text{e}}(\ell )$  \eqref{eq:DefTauEddies} of eddies of typical size $\ell$ would then follow a power-law with exponent $2/3$ as a function of the length scale $\ell $. One may wonder whether this prediction is correct. The proposed definition of the time scale $\tau_{\text{e}}(\ell )$  \eqref{eq:DefTauEddies} clearly depends solely on the spatial structure of the velocity field and is independent of the temporal structure of the velocity field. In order to take into account the temporal structure, let us thus define a characteristic time scale $\tau_c(\ell )$ which is based on the correlation in time of the spatial increment given by
\begin{align}\label{eq:DefTauEddiesCorrTime}
\tau_{\text{c}}(\ell ) \equiv \frac{1}{S_2(\ell)} \lim_{\nu\to 0} \int_{0}^{\infty}\E \left[\delta_\ell \bu^\nu(t,\bx)\cdot \delta_\ell \bu^\nu(t+\tau,\bx) \right]\rd \tau.
\end{align}
As it will be argued below, we could also define a similar time scale as $\tau_{\text{c}}(\ell )$ \eqref{eq:DefTauEddiesCorrTime} to characterize the turnover time scale of an eddy of size $\ell$ using instead the temporal structure of the velocity Fourier modes at a given wave vector $k$. This indicates that one {associates} in a loose sense the length scale $\ell $ to $|\bk|^{-1}$. This must be treated with great care since we have assumed in this Chapter that the spatial domain is infinite, i.e. $\bx\in\R^3$, and thus Fourier modes must be considered as distributions. Nevertheless, we would define in a formal way the following time scale 
\begin{align}\label{eq:DefTauF}
\tau_{\text{f}}(k) \equiv ``\lim_{\nu\to 0} \frac{1}{\E \left[\left|\widehat{\bu}^\nu(t,\bk)\right|^2\right]}\int_{0}^{\infty}\E \left[\widehat{\bu}^\nu(t,\bk)\cdot \overline{\widehat{\bu}^\nu(t+\tau,\bk)} \right]\rd \tau ",
\end{align}
where $\widehat{\bu}^\nu(t,\bk)$ is the Fourier transform of the velocity field $\bu^\nu(t,\bx)$, and $\overline{\cdot }$ standing for the complex conjugation. The definition of the time scale $\tau_{\text{f}}(k)$ \eqref{eq:DefTauF} is for the time being put in quotes because, as we will see, the Fourier transform over all space
has a distributional nature, and necessitates the formulation of the model in a periodic setting. 

 Note that the time scale $\tau_{\text{f}}(k)$ is expected to depend solely on the norm $k$ of the wave vector $\bk$ by statistical isotropy and can be interpreted as the typical correlation duration at a given wave number. The two time scales $\tau_{\text{c}}(\ell )$ \eqref{eq:DefTauEddiesCorrTime} and $\tau_{\text{f}}(k)$ \eqref{eq:DefTauF} are related according to
\begin{align}\label{eq:RelTauCTauF}
\tau_{\text{c}}(\ell ) =``\frac{\int_{\bk\in\R^3}\left[ 1-\cos\left( 2\pi \bk \cdot \bell\right)\right]\tau_{\text{f}}(k)\E \left[\left|\widehat{\bu}(t,\bk)\right|^2\right]\rd\bk}{\int_{\bk\in\R^3}\left[ 1-\cos\left( 2\pi \bk \cdot \bell\right)\right]\E\left[ \left|\widehat{\bu}(t,\bk)\right|^2\right]\rd\bk}",
\end{align}
where we have denoted $\E\left[ \left|\widehat{\bu}(t,\bk)\right|^2\right]$ the limit  as $\nu\to 0$ of the PSD $\E\left[ \left|\widehat{\bu}^\nu(t,\bk)\right|^2\right]$. Once again, we recall that expressions such as \eqref{eq:RelTauCTauF} should be taken in a formal way. Only the consideration of periodic boundary conditions gives a proper meaning to such quantities.

With its definition \eqref{eq:DefTauF}, the characteristic time scale $\tau_{\text{f}}(k)$ has been estimated in DNS by several groups \citep{KanIsh99,FavGod10,GorBal21} while assuming periodic boundary conditions instead of the full space $\R^3$, leading to  well defined Fourier modes. The numerical investigations from the articles quoted above show, in a consistent manner with the dimensional analysis of \cite{Ten75}, that
\begin{align}\label{eq:behaviorTauFLargeK}
\tau_{\text{f}}(k) \build{\propto}_{k\to\infty}^{} \frac{1}{\sigma k}.
\end{align}
The behavior of the time scale $\tau_{\text{f}}(k)$ \eqref{eq:behaviorTauFLargeK} at large wave numbers as $k^{-1}$ suggests, based on the dimensional ground, that the typical velocity entering in the motion of eddies of size $k^{-1}$ is the standard deviation of the velocity $\sigma$, i.e. the characteristic velocity of large scales. To see this more clearly, let us see the implications of this behavior on the time scale $\tau_{\text{c}}(\ell )$ \eqref{eq:DefTauEddiesCorrTime}. To do so, consider the identity displayed in \eqref{eq:RelTauCTauF} and rescale the dummy integration variable $\bk$ by $\ell $, then one obtains
\begin{align}\label{eq:RelTauCTauFRescale}
\tau_{\text{c}}(\ell ) &=``\frac{\int_{\bk\in\R^3}\left[ 1-\cos\left( 2\pi \bk \cdot \frac{\bell}{\ell }\right)\right]\tau_{\text{f}}(|\bk|/\ell )\E \left[\left|\widehat{\bu}(t,\bk/\ell )\right|^2\right]\rd\bk}{\int_{\bk\in\R^3}\left[ 1-\cos\left( 2\pi \bk \cdot \frac{\bell}{\ell }\right)\right]\E\left[ \left|\widehat{\bu}(t,\bk/\ell )\right|^2\right]\rd\bk}"\notag\\
&\build{\propto}_{\ell \to 0}^{}\frac{\ell }{\sigma}\frac{\int_{\rho\ge 0,\theta\in[0,\pi]}\left[ 1-\cos\left( 2\pi \rho\cos \theta\right)\right]\rho^{-8/3}\rd\rho \rd\theta}{\int_{\rho\ge 0,\theta\in[0,\pi]}\left[ 1-\cos\left( 2\pi \rho\cos \theta\right)\right]\rho^{-5/3}\rd\rho \rd\theta},
\end{align}
where we have used the asymptotic behaviors of $\tau_{\text{f}}(k)$ \eqref{eq:behaviorTauFLargeK} and of the PSD $\E \left[\left|\widehat{\bu}(t,\bk)\right|^2\right]\sim |\bk|^{-11/3}$ \eqref{eq:DeducedModelPSD3d} at large argument. Notice that the remaining ratio of double integrals entering in \eqref{eq:RelTauCTauFRescale} can be shown to be finite (an exact expression can even be obtained using a symbolic calculation software). The relevant result given in \eqref{eq:RelTauCTauFRescale} is the fact that, as a consequence of the observation that $\tau_{\text{f}}(k)$ becomes inversely proportional to $k$ at large $k$ \eqref{eq:behaviorTauFLargeK}, the genuine characteristic time scale $\tau_{\text{c}}(\ell )$ of eddies of typical size $\ell $ \eqref{eq:DefTauEddiesCorrTime} becomes proportional to $\ell $ \eqref{eq:RelTauCTauFRescale}. Consequently, the relevant characteristic velocity appearing in \eqref{eq:RelTauCTauFRescale} is $\sigma$, i.e., the characteristic velocity of the largest eddies, while it may have been more natural to obtain their typical velocity $\sqrt{S_2(\ell)}$ as in the expression \eqref{eq:DefTauEddies} of $\tau_{\text{e}}(\ell )$: this is the phenomenon of advection of the small scales by the large eddies, referred to here and in the turbulence literature as the sweeping effect.

\subsection{Implications for the spatio-temporal covariance matrix of the velocity field}\label{Par:PresentModelOnR3}

Let us now present the general form of the covariance structure in time and space of the velocity field that encompasses both the spatial structure detailed in Paragraph \ref{Sec:SpatialStats} and the peculiar temporal structure which is governed by the sweeping effect, as is in the former discussion \ref{Par:DiscussionSweeping}.
To do so, let us begin by defining the relevant spatio-temporal covariance of the velocity field $\mathcal C_{ij}^{\nu}(\tau,\bell)$ as
\begin{align}\label{eq:DefSpatioTempCorr}
\mathcal C_{ij}^{\nu}(\tau,\bell) = \E \left[ u_i^{\nu}(t,\bx)u_j^{\nu}(t+\tau,\bx+\bell)\right].
\end{align}
Again, $\mathcal C_{ij}^{\nu}(\tau,\bell)$ is independent of position $\bx$ and time $t$ by statistical homogeneity and stationarity, respectively. We mention that, moreover, the spatiotemporal covariance function \eqref{eq:DefSpatioTempCorr} is an even function of the temporal argument $\tau\in\R$, $\mathcal C_{ij}^{\nu}(-\tau,\bell) = \mathcal C_{ij}^{\nu}(\tau,\bell)$. In the following, we will also assume statistical isotropy and explore the consequences of this assumption on the structure of the covariance function. The natural extension of the spatial covariance matrix \eqref{eq:FullSpatCov} to a spatio-temporal framework $\mathcal C_{ij}^{\nu}(\tau,\bell)$ \eqref{eq:DefSpatioTempCorr} can be conveniently written in Fourier space as 
\begin{align}\label{eq:SpatioTempCorrFour}
\mathcal C_{ij}^{\nu}(\tau,\bell) = \frac{1}{2} \int_{\bk\in \R^3} e^{2i\pi \bk \cdot \bell}F(\tau/T_\bk) E_\nu^{\text{\tiny{E}}}(\bk) \widehat{P}_{ij}(\bk )\ \rd \bk ,
\end{align}
where $F$ encodes the temporal dependence, assumed to be an integrable function normalized such that $F(0)=1$, $T_\bk$ a time scale that depends on the wave vector $\bk$, $ E_\nu^{\text{\tiny{E}}}(\bk) $ is the PSD of the velocity field \eqref{eq:DefPSD} studied above, and $\widehat{P}_{ij}(\bk )$ the Leray projector on divergence-free vector fields \eqref{eq:Leray}. The covariance structure \eqref{eq:SpatioTempCorrFour} was initially proposed by \cite{ChaGaw03} when $F$ is chosen such that $F(\tau)=\exp(-|\tau|)$. As it will be explained below, this choice leads to a  Markovian evolution for the Fourier modes, as early noticed by \cite{KomPes04}.

As we have already mentioned, it is clear from \eqref{eq:SpatioTempCorrFour} that the PSD $E_\nu^{\text{\tiny{E}}}(k)$ corresponds to the Fourier transform of $\mathcal C_{ii}^{\nu}(0,\ell)$. As a consequence, we recover from \eqref{eq:SpatioTempCorrFour} the scaling behaviors of the second-order structure functions \eqref{eq:DefS2Holder13LongWithCk} and \eqref{eq:DefS2Holder13}. To make a connection with the discussion developed in Section \ref{Par:DiscussionSweeping}, let us establish the link between the characteristic time scale $\tau_{\text{c}}(\ell )$ \eqref{eq:DefTauEddiesCorrTime} and the $k$-dependent time scale $T_k$ entering in $\mathcal C_{ij}^{\nu}(\tau,\bell)$ \eqref{eq:SpatioTempCorrFour}. We have
\begin{align}\label{eq:RelatTauCTk}
\tau_{\text{c}}(\ell ) &= \frac{\int_{\tau=0}^{\infty}\left[\mathcal C_{ii}^{\nu}(\tau,0)-\mathcal C_{ii}^{\nu}(\tau,\bell)\right]\rd \tau}{\mathcal C_{ii}^{\nu}(0,0)-\mathcal C_{ii}^{\nu}(0,\bell)}\notag\\
&=\frac{\int_{\bk \in \R^3} \left(1-e^{2i\pi \bk \cdot \bell} \right)T_\bk E^{\text{\tiny{E}}}(|\bk|)\rd \bk}{\int_{\bk \in \R^3} \left(1-e^{2i\pi \bk \cdot \bell} \right) E^{\text{\tiny{E}}}(|\bk|)\rd \bk}\int_0^{\infty} F(s)\rd s.
\end{align}
As suggested, in a formal manner, by the dependence on the wave number $k$ of the characteristic time scale $\tau_{\text{f}}(k)$ \eqref{eq:DefTauF}, let us similarly assume, as has been done by \cite{ChaGaw03}, that 
\begin{align}\label{eq:DependTkOnKLargeK}
T_k\build{\sim}_{k\to\infty}^{}\frac{1}{D_3 k},
\end{align}
where $D_3$ is a free parameter of the description that has the dimension of a velocity. Similarly to the derivation of the relationship between $\tau_{\text{c}}(\ell ) $ and $\tau_{\text{f}}(k)$ obtained in \eqref{eq:RelTauCTauFRescale}, using the scaling behavior obtained in \eqref{eq:DependTkOnKLargeK} and the behavior of the PSD in the same limit \eqref{eq:DeducedModelPSD3d}, we get from \eqref{eq:RelatTauCTk}, while performing the remaining ratio of integrals expressed in spherical variables,
\begin{align}\label{eq:RelatTauCTkAsympt}
\tau_{\text{c}}(\ell ) &\build{\sim}_{\ell \to 0}^{}\frac{\ell }{D_3}\frac{2^{1/3} \sqrt{\pi} \Gamma^2(1/6)}{15 \Gamma(11/6)}\int_0^{\infty} F(s)\rd s.
\end{align}
The asymptotic behavior   in \eqref{eq:RelatTauCTkAsympt}, assuming the asymptotic behavior of $T_k$ \eqref{eq:DependTkOnKLargeK} is consistent with the expected behavior given in \eqref{eq:RelTauCTauFRescale}. This suggests that the free parameter $D_3$ can be expressed in units of the velocity standard-deviation $\sigma$ \eqref{eq:AsymptVariance}.

Although physically insightful, the behavior of the characteristic time scale $\tau_{\text{c}}(\ell )$ with the eddy's size $\ell $ \eqref{eq:RelatTauCTkAsympt} is not traditionally estimated in experiments and numerical simulations. Instead, the temporal second-order structure function and the time spectrum are usually studied. 

Concerning the temporal second-order structure function $S_2^{\text{\tiny{T}}}(\tau)$, let us define the velocity time increment, i.e. the variation of velocity in time at a fixed position $\bx$,
\begin{align}\label{eq:TimeIncrement}
\delta_\tau \bu^\nu(t,\bx)=\bu^\nu(t+\tau,\bx)-\bu^\nu(t,\bx).
\end{align}
Using the covariance structure of the velocity field \eqref{eq:SpatioTempCorrFour}, we obtain
\begin{align}\label{eq:TemporalS2}
S_2^{\text{\tiny{T}}}(\tau)\equiv \lim_{\nu\to 0}\E\left[ \left|\delta_\tau \bu^\nu\right|^2\right]&=2\left[ \mathcal C_{ii}(0,0) -\mathcal C_{ii}(\tau,0) \right]\notag\\
&=2\int_{\bk \in \R^3} \left[1-F(\tau/T_\bk) \right]E^{\text{\tiny{E}}}(|\bk|)\rd \bk\notag\\
&= 2\int_{\bk \in \R^3} \left[1-F(\tau/T_{\bk/\tau}) \right]E^{\text{\tiny{E}}}(|\bk|/\tau)\rd \bk/\tau^3,
\end{align}
such that, in the limit of vanishing time scales $\tau\to 0^+$, using the  asymptotic behaviors of $T_\bk$ \eqref{eq:DependTkOnKLargeK} and $E^{\text{\tiny{E}}}(|\bk|)$ \eqref{eq:DeducedModelPSD3d} at large wave vector amplitude, one obtains
\begin{align}\label{eq:TemporalS2LimTau0}
S_2^{\text{\tiny{T}}}(\tau)&\build{\sim}_{\tau\to 0^+}^{} 4\frac{55}{9}D_2 D_3^{2/3}\tau^{2/3} \int_{0}^{\infty} \left[1-F(\rho) \right]\rho^{-5/3} \rd\rho.
\end{align}
The integral in the right-hand side of  \eqref{eq:TemporalS2LimTau0} is finite as long as $1-F(\rho)$ goes to 0 as $\rho^a$ with $a>2/3$, which will be the case in all future situations. Expressing $D_2$ in units of the average dissipation $\varepsilon$ (compare, for instance, the expressions provided in \eqref{eq:DefS2Holder13Long} and \eqref{eq:DefS2Holder13LongWithCk}) and $D_3$ in units of velocity standard-deviation $\sigma$, the scaling behavior obtained for the temporal second-order structure function \eqref{eq:TemporalS2LimTau0} can be alternatively written as 
\begin{align}\label{eq:PredVarTimeIncrement}
S_2^{\text{\tiny{T}}}(\tau)&\build{\propto}_{\tau\to 0^+}^{} \sigma ^{2/3} (\varepsilon \tau)^{2/3},
\end{align}
which coincides with the dimensional prediction of \cite{Ten75}.

Similarly, let us consider the corresponding temporal (or frequency) spectrum $E_\nu^{\text{\tiny{T}}}(\omega)$ defined by
\begin{align}\label{eq:DefTimeSpectrum}
E_\nu^{\text{\tiny{T}}}(\omega) &= \int_{\tau\in\R}e^{-2i\pi \omega \tau} \E \left[ \bu^{\nu}(t,\bx)\cdot \bu^{\nu}(t+\tau,\bx)\right]\rd \tau = \int_{\tau\in\R}e^{-2i\pi \omega \tau} \mathcal C_{ii}^{\nu}(\tau,0)\rd \tau\\
&= \int_{\bk \in \R^3} \int_{\tau\in\R} e^{-2i\pi \omega \tau} F(\tau/T_\bk) E_\nu^{\text{\tiny{E}}}(\bk)\rd \bk\ \rd \tau\notag\\
&= \int_{\bk \in \R^3} T_\bk\widehat{F}(\omega T_\bk) E_\nu^{\text{\tiny{E}}}(\bk)\rd \bk\notag\\
&= \int_{\bk \in \R^3} T_{\omega \bk}\widehat{F}(\omega T_{\omega \bk}) E_\nu^{\text{\tiny{E}}}(\omega |\bk|) \omega^3 \ \rd \bk ,\label{eq:ExpressionTimeSpectrum}
\end{align}
where $\widehat{F}$ is the Fourier transform of the function $F$. Once again, using the expressions of $T_\bk$ \eqref{eq:DependTkOnKLargeK} and $E^{\text{\tiny{E}}}(|\bk|)$ \eqref{eq:DeducedModelPSD3d} at large wave vector amplitude, we obtain
\begin{align}\label{eq:TimeSpectrumAsympt}
\lim_{\nu\to 0}E_\nu^{\text{\tiny{T}}}(\omega) &= \int_{\bk \in \R^3} \omega T_{\omega \bk}\widehat{F}(\omega T_{\omega \bk}) E^{\text{\tiny{E}}}(\omega |\bk|) \omega^2\rd\bk\notag\\
&\build{\sim}_{\omega\to\infty}^{}2\frac{55}{9}D_2D_3^{2/3}\omega^{-5/3}\int_{0}^{\infty} \widehat{F}\left(\rho^{-1}\right) \rho^{-8/3} \rd\rho.
\end{align}
The integral in the right-hand side of \eqref{eq:TimeSpectrumAsympt} will be shown to be finite for the classes of functions $F$ considered below. After being reformulated in units of $\varepsilon$ and $\sigma$, similarly to what has been done in \eqref{eq:PredVarTimeIncrement}, the frequency dependence of the time spectrum \eqref{eq:TimeSpectrumAsympt} is given by
\begin{align}\label{eq:TimeSpectrumAsymptEpsilonSigma}
\lim_{\nu\to 0}E_\nu^{\text{\tiny{T}}}(\omega) \build{\propto}_{\omega\to \infty}^{} \sigma ^{2/3} \varepsilon^{2/3}\omega^{-5/3},
\end{align}
which is consistent with the dimensional predictions of \cite{Ten75}.

\section{Gaussian random velocity vector, and its {Markovian}  evolution}\label{Sec:FormModel}

\subsection{Formulation of the model with periodic boundary conditions: statistical structure of Fourier modes}\label{Sec:FormModelIntroPerio}

For practical and numerical reasons, we   formulate a version of the Gaussian model described in the former section on the  {three dimensional periodic space $[-\Ltot/2\ ;\ \Ltot/2]^3$ of period} $\Ltot$, as has been done in DNS. The model is Gaussian, it is thus fully determined in the statistically homogeneous and stationary regime by its covariance function \eqref{eq:DefSpatioTempCorr}.
 The respective velocity field  will thus thus $L_{\text{\tiny{tot}}}$-periodic $u^\nu_i(t,\bx)=u^\nu_i(t,\bx+L_{\text{\tiny{tot}}}\hat{e}_p)$ in any Cartesian direction $p=1$, $2$ or $3$. Let us mention that future simulations will be performed using $L_{\text{\tiny{tot}}}= 2\pi$ in order to be consistent with the DNS. {A fundamental consequence of considering a periodic setting   is the possibility to expand the velocity field using a Fourier series and the respective Fourier modes $\widehat{u}^\nu_i(t,\bk_\bn)$, depending on $\bk_\bn = \bn/L_{\text{\tiny{tot}}}$ for  $\bn=(n_1,n_2,n_3)\in\Z^3$.}
It can {then} be checked that the variance of the Fourier modes is finite. This solves the issue exhibited in the former Section, where Fourier modes in the whole space need to be interpreted in a distributional sense. {Since the velocity field $\bu^\nu$ is real-valued, Fourier modes are hermitian symmetric, i.e. $\overline{\widehat{u}^\nu_i(t,\bk_\bn)} = \widehat{u}^\nu_i(t,-\bk_\bn)= \widehat{u}^\nu_i(t,\bk_{-\bn})$.} This implies that the Fourier mode for $\bn=(0,0,0)$ is real-valued.

The purpose of this Section is the formulation in a periodic setting of the statistical quantities defined in the whole space in Section \ref{Par:PresentModelOnR3}. To do so, let us begin by defining the   spatio-temporal covariance of the velocity field $\mathcal C_{ij}^{\nu}(\tau,\bell)$ as
\begin{align}\label{eq:DefSpatioTempCorrPeriodic}
\mathcal C_{ij}^{\nu}(\tau,\bell) = \E \left[ u_i^{\nu}(t,\bx)u_j^{\nu}(t+\tau,\bx+\bell)\right].
\end{align}
The definition is identical to which is very similar to \eqref{eq:DefSpatioTempCorr}, however $\mathcal C_{ij}^{\nu}(\tau,\bell) $ given in \eqref{eq:DefSpatioTempCorrPeriodic} is periodic. If it is furthermore required that $\mathcal C_{ij}^{\nu}$ \eqref{eq:DefSpatioTempCorrPeriodic} characterizes a statistically homogeneous and stationary incompressible velocity field, the right-hand side of \eqref{eq:DefSpatioTempCorrPeriodic} is thus  independent on the position $\bx$ and time $t$. The covariance structure of the Fourier modes is then written as
\begin{align}\label{eq:CovFourierModesPerio}
\E \left[ \widehat{u}_i^{\nu}(t,\bk_\bn)\widehat{u}_j^{\nu}(t+\tau,\bk_\bm)\right] = \frac{L^3_{\text{\tiny{tot}}}}{2}F(\tau/T_{k_\bn}) E_\nu^{\text{\tiny{E}}}(k_\bn)\widehat{P}_{ij}(\bk_\bn)\delta_{\bn,-\bm}^{(3)},
\end{align}
where $\widehat{P}_{ij}(\bk_\bn)$ is the Leray projector \eqref{eq:Leray} at the discrete wave vector $\bk_\bn$ (whose norm is $k_\bn$ and also corresponds to $k_n$ in virtue of isotropy), $\delta_{\bn,\bm}^{(3)}=\prod_i\delta_{n_i,m_i}$ the three-dimensional Kronecker delta function over the indices $\bn$ and $\bm$, scalar functions $F$ and $T_{k_\bn}$ encoding the temporal structure similarly to what we have seen in Section \ref{Par:PresentModelOnR3}, and the respective PSD defined as the expectation of the norm square of the Fourier mode, i.e.
\begin{align}\label{eq:RelPSDVarFourierModesPerio}
\E \left[\left|\widehat{\bu}^{\nu}(t,\bk_\bn)\right|^2\right]= L^3_{\text{\tiny{tot}}}E_\nu^{\text{\tiny{E}}}(k_\bn).
\end{align}
This allows to write the covariance of the velocity field \eqref{eq:DefSpatioTempCorrPeriodic} as
\begin{align}\label{eq:SpatioTempCorrPeriodicFourierSeries}
\mathcal C_{ij}^{\nu}(\tau,\bell) = \frac{1}{2L^3_{\text{\tiny{tot}}}}\sum_{ \bn\in\Z^3} e^{2i\pi\bk_\bn\cdot \bell}F(\tau/T_{k_\bn}) E_\nu^{\text{\tiny{E}}}(k_\bn)\widehat{P}_{ij}(\bk_\bn).
\end{align}

For the sake of generality, let us consider an expression of the uni-dimensional PSD $E_\nu^{\text{\tiny{E,long}}}(k)$ \eqref{eq:ModelPSD1dLong}, which we recall to deal asymptotically with H\"older-continuous functions of parameter $H=1/3$, as observed in fluid turbulence, to any parameter $0<H<1$. For these regularities, the expected PSD would also exhibit a power-law behavior as in \eqref{eq:ModelPSD1dLong}, but with an exponent that would depend on $H$, of the form
\begin{align}\label{eq:ModelPSD1dLongAnyH}
E_\nu^{\text{\tiny{E,long}}}(k) =D_2 |k|_{1/L}^{-(2H+1)} e^{-\eta_d k},
\end{align}
where the dissipative cut-off $\eta_d$ {is a parameter of the formulation. Let us recall that the usual phenomenology of turbulence, as depicted in a self-consistent manner in Section \ref{Sec:SpatialStats} is recovered only for $H=1/3$.}  In the following, we will consider the length scale $\eta_d$ as a free parameter. As we have seen in  Section \ref{Sec:SpatialStats}, imposing a given PSD in the longitudinal case $E_\nu^{\text{\tiny{E,long}}}(k)$ \eqref{eq:ModelPSD1dLong} implies for statistical isotropic reasons a form of the PSD $E_\nu^{\text{\tiny{E}}}(k)$ in dimension $d=3$ through the relation provided in \eqref{eq:RelPSD1dLong3d}, giving a particular simple form in the asymptotic limit of vanishing viscosity $E^{\text{\tiny{E}}}(k)$ \eqref{eq:DeducedModelPSD3d}. Using any $0<H<1$, instead of the particular value $H=1/3$, the one-dimensional spectrum \eqref{eq:ModelPSD1dLongAnyH} leads to, through the relation \eqref{eq:RelPSD1dLong3d},
\begin{align}\label{eq:ModelPSDAnyHNuTo0}
\lim_{\nu\to 0}E_\nu^{\text{\tiny{E}}}(k) = \frac{1}{2\pi}(1+2H)(3+2H) D_2 k^2 |k|_{1/L}^{-(2H+5)},
\end{align}
which is consistent with \eqref{eq:DeducedModelPSD3d} for $H=1/3$. Also, for the sake of generality, and to make a connection with \cite{ChaGaw03}, the time scale $T_k$ will be chosen to be 
\begin{align}\label{eq:DefTkBeta}
T_k = \frac{1}{D_3 |k|_{1/L}^{2\beta}},
\end{align}
where $\beta$ is an additional new free parameter of the model, which takes the particular value $\beta=1/2$ to ensure the asymptotic behavior observed in turbulence \eqref{eq:DependTkOnKLargeK}. The remaining free parameter $D_3$ is expected to be proportional to the velocity standard deviation only in the case $\beta =1/2$ and $H=1/3$. 

In the limit of an infinitely large periodic box $L_{\text{\tiny{tot}}}\to\infty$, we obtain the correspondence for any appropriate function $f$
\begin{align}\label{eq:CorrespLtotInfinite}
\lim_{L_{\text{\tiny{tot}}}\to\infty} \frac{1}{L^3_{\text{\tiny{tot}}}}\sum_{ \bn\in\Z^3} f(\bk_\bn)=\int_{\bk\in\R^3}f(\bk)\ \rd \bk ,
\end{align}
such that in particular the velocity covariance function \eqref{eq:SpatioTempCorrPeriodicFourierSeries} tends to its continuous counterpart \eqref{eq:SpatioTempCorrFour}. As a consequence, all relevant statistical quantities such as variance predictions \eqref{eq:VarianceFromPSD} and average dissipation \eqref{eq:AverageDissFromPSD}
remain valid in the periodic setting for $H=1/3$, and can easily be generalized for any $0<H<1$. 

Let us also explore the statistical structure of the spatial structure functions at small scales implied by the power-law behavior of the PSD given in \eqref{eq:ModelPSD1dLongAnyH}. Concerning the variance of the longitudinal velocity increment \eqref{eq:DefVelIncrVectLong},  instead of \eqref{eq:DefS2Holder13Long}, we obtain
\begin{align}\label{eq:DefS2HolderHLong}
S_2^{\text{\tiny{long}}}(\ell) &= \lim_{L_{\text{\tiny{tot}}}\to\infty} \lim_{\nu\to 0} \E \left[ \left(\delta_\ell^{\text{\tiny{long}}} \bu^\nu\right)^2\right]= 2\int_{k\in\R} \left( 1-e^{2i\pi k\ell }\right) E^{\text{\tiny{E,long}}}(k) \rd k\notag\\
&\build{\sim}_{\ell \to 0}^{} 2D_2(2 \pi)^{2 H} \cos(\pi H) \frac{\Gamma(2-2 H)}{H(1-2H)}\ell ^{2H}.
\end{align}
Using the well-known relation $\Gamma(z+1)=z\Gamma(z)$, one checks that when $H=1/3$, the prediction \eqref{eq:DefS2HolderHLong} coincides  with \eqref{eq:DefS2Holder13Long}. Concerning the increment of a   velocity component \eqref{eq:DefVelocityIncrementCompI} and the   second-order structure function \eqref{eq:DefSijNuTo0},
we derive under the assumption of statistical isotropy and incompressibility a generalization of \eqref{eq:SijIsotropyIncomp}{, one has :}
\begin{align}\label{eq:SijIsotropyIncompAnyH}
S_{2,ij}(\bell) \build{\sim}_{\ell \to 0}^{} S_2^{\text{\tiny{long}}}(\ell )\left( -H\frac{\ell_i\ell_j}{\ell ^2}+(1+H)\delta_{ij}\right),
\end{align}
with the consequence that 
\begin{align}\label{eq:DefS2HolderAnyH}
S_2(\ell)\equiv S_{2,ii}(\ell) = \lim_{L_{\text{\tiny{tot}}}\to\infty} \lim_{\nu\to 0} \E \left[ \left|\delta_\ell \bu^\nu\right|^2\right] &= 2\int_{\bk\in\R^3} \left( 1-e^{2i\pi \bk \cdot \bell}\right) E^{\text{\tiny{E}}}(|\bk|)\rd \bk\notag\\
&\build{\sim}_{\ell \to 0}^{}(3+2H)S_2^{\text{\tiny{long}}}(\ell ).
\end{align}

Correspondingly, the characteristic time scale $\tau_{\text{c}}(\ell )$ \eqref{eq:DefTauEddiesCorrTime} reads, using the expression provided in \eqref{eq:RelatTauCTk} with the generalized forms of $ E^{\text{\tiny{E}}}(|\bk|)$ \eqref{eq:ModelPSDAnyHNuTo0} and $T_\bk$ \eqref{eq:DefTkBeta}, and for any appropriate function $F$, in the limit of an infinitely large periodic box allowing to use the {correspondence} depicted in \eqref{eq:CorrespLtotInfinite}, 
\begin{align}\label{eq:RelatTauCTkLtotInfinite}
\lim_{L_{\text{\tiny{tot}}}\to\infty}\tau_{\text{c}}(\ell ) &=\frac{\int_{\bk \in \R^3} \left(1-e^{2i\pi \bk \cdot \bell} \right)T_\bk E^{\text{\tiny{E}}}(|\bk|)\rd\bk}{\int_{\bk \in \R^3} \left(1-e^{2i\pi \bk \cdot \bell} \right) E^{\text{\tiny{E}}}(|\bk|)\rd\bk}\int_{0}^\infty F(s)\rd s.
\end{align}
At small scales, the characteristic time scale $\tau_{\text{c}}(\ell )$ \eqref{eq:RelatTauCTkLtotInfinite} undergoes a transition depending on the values of $H$ and $\beta$, following the behaviors
\begin{align}\label{eq:RelatTauCTkLtotInfiniteTrans1}
\lim_{L_{\text{\tiny{tot}}}\to\infty}\tau_{\text{c}}(\ell ) \build{\sim}_{\ell \to 0}^{}\frac{\ell ^{2\beta}}{D_3}\frac{\int_{\rho\ge0} \left(1-\frac{\sin(2\pi \rho)}{2\pi \rho} \right)\rho^{-2H-1-2\beta}\rd \rho}{\int_{\rho\ge0} \left(1-\frac{\sin(2\pi \rho)}{2\pi \rho} \right)\rho^{-2H-1}\rd\rho}\int_{0}^\infty F(s)\rd s \mbox{ for } H<1-\beta,
\end{align}
and
\begin{align}\label{eq:RelatTauCTkLtotInfiniteTrans2}
\lim_{L_{\text{\tiny{tot}}}\to\infty}&\tau_{\text{c}}(\ell ) \notag\\
&\build{\sim}_{\ell \to 0}^{}\frac{4\pi^2}{3}\frac{\ell ^{2(1-H)}}{D_3}\frac{\int_{\rho\ge0} \rho^6\sqrt{\rho^2+1/L^2}^{-2H-5-2\beta}\rd \rho}{\int_{\rho\ge0} \left(1-\frac{\sin(2\pi \rho)}{2\pi \rho} \right)\rho^{-2H-1}\rd \rho}\int_{0}^\infty F(s)\rd s \mbox{ for } H>1-\beta.
\end{align}
Expressions provided in \eqref{eq:RelatTauCTkLtotInfiniteTrans1} and \eqref{eq:RelatTauCTkLtotInfiniteTrans2} could be further written in an explicit way with the help of a symbolic calculation software.

\subsection{Markovian evolution of the Fourier modes}\label{Sec:CausalEvolFM}
Let us now discuss the time evolution of the Fourier modes $\widehat{u}^\nu_i(t,\bk_\bn)$, such that the correlation structure is given by \eqref{eq:CovFourierModesPerio}, leading to the expected covariance structure $\mathcal C_{ij}^{\nu}(\tau,\bell) $ given in \eqref{eq:SpatioTempCorrPeriodicFourierSeries}. This cannot be done in full generality for any function $F$ since we are asking for a Markovian evolution. Actually, very few choices for $F$ can be achieved for a given {Markovian} evolution, and we will only explore a very limited set of them. As far as we are concerned, we will present choices of $F$ which give a {Markovian} evolution {in which additional underlying layers will be involved}.

{Before proceeding, we have to define a space-periodic and temporal Gaussian white noise ``$\rd W_i(t,\bx)"$, a notation that we put in quotes because it has to be considered as a distribution. More precisely, let us focus on its Fourier modes ``$\widehat{\rd W}_i(t,\bk_\bn)"$. Being complex Gaussian and Hermitian, it is eventually fully determined by the covariance function 
\begin{align}\label{eq:CovFourierModesWSpaceTime}
\E\left[\widehat{\rd W}_i(t,\bk_\bn)\widehat{\rd W}_j(t',\bk_\bm) \right]=L^3_{\text{\tiny{tot}}}\delta_{i,j}\delta_{\bn,-\bm}^{(3)}\delta(t-t')\ \rd t\ \rd t',
\end{align}
where $\delta(t)$ stands for the Dirac-delta distribution.}

\subsubsection{The statistical structure of the stationary solution: spatial Fractional Gaussian vector Field}\label{Sec:DefFGF}
 The stochastic evolutions that are proposed below are  designed such that their statistically stationary solutions are well defined, of finite variance and all coincide with a so-called fractional Gaussian vector (fGv) field, that we will denote $\bu^{\text{\tiny{fGv}},\nu}(\bx)$. The corresponding Fourier modes are denoted by $\widehat{\bu}^{\text{\tiny{fGv}},\nu}(\bk_\bn)$.

{Let us first consider the purely spatial white noise ``$\rd \bW(\bx)$'', with (Hermitian symmetric) Fourier modes $\widehat{\rd \bW}(\bk_\bn)$ covariance given by
\begin{align}\label{eq:CovFourierModesWSpace}
\E\left[\widehat{\rd W}_i(\bk_\bn)\widehat{\rd W}_j(\bk_\bm) \right]=L^3_{\text{\tiny{tot}}}\delta_{i,j}\delta_{\bn,-\bm}^{(3)}.
\end{align}
The real random field $\bu^{\text{\tiny{fGv}},\nu}(\bx)$ is fully defined by its Fourier mode, and is given, component-wise, by
\begin{align}\label{eq:DefFMfGv}
\widehat{u}^{\text{\tiny{fGv}},\nu}_i(\bk_\bn) = \sqrt{\frac{E_\nu^{\text{\tiny{E}}}(k_\bn)}{2}}\widehat{P}_{ip}(\bk_\bn)\widehat{\rd W}_p(\bk_\bn).
\end{align}
The covariance coincides with \eqref{eq:CovFourierModesPerio} when $\tau=0$, i.e.
\begin{align}\label{eq:CovFourierModesPeriofGv}
\E \left[\widehat{u}^{\text{\tiny{fGv}},\nu}_i(\bk_\bn)\widehat{u}^{\text{\tiny{fGv}},\nu}_j(\bk_\bm)\right] = \frac{L^3_{\text{\tiny{tot}}}}{2} E_\nu^{\text{\tiny{E}}}(k_\bn)\widehat{P}_{ij}(\bk_\bn)\delta_{\bn,-\bm}^{(3)}.
\end{align}
}

\subsubsection{The Markovian evolution of \cite{ChaGaw03}}\label{Sec:DefChavesGawOneLayer}

The first natural evolution for Fourier modes, which coincides with the proposition of \cite{ChaGaw03}, as early recognized by \cite{KomPes04}, is of Ornstein-Uhlenbeck type, genuinely Markovian, and reads as the stochastic differential equation
\begin{equation}\label{eq:GeneralOUTkFourierModes}
d\widehat{u}^{(1),\nu}_i(t,\bk) = -\frac{1}{T_{\bk}}\widehat{u}^{(1),\nu}_i(t,\bk)\rd t+\sqrt{\frac{E_\nu^{\text{\tiny{E}}}(\bk)}{T_{\bk}}}\widehat{P}_{ip}(\bk)\widehat{\rd W}_p(t,\bk),
\end{equation}
for a given discrete wave vector $\bk=\bk_\bn$.  $T_\bk$ is the characteristic time of the Fourier mode $\bk$ defined in \eqref{eq:DefTkBeta}, $\widehat{P}_{ij}(\bk )$ Leray projector ensuring the incompressibility defined in \eqref{eq:Leray} and  $E_\nu^{\text{\tiny{E}}}(k)$ {the} previously defined 3D-PSD obtained from \eqref{eq:ModelPSD1dLongAnyH} using the isotropic relation \eqref{eq:RelPSD1dLong3d}. The superscript $^{(1)}$ appearing in the notation of the resulting velocity field $\widehat{u}^{(1),\nu}_i(t,\bk)$ \eqref{eq:GeneralOUTkFourierModes} illustrates that it is obtained using a single evolution with respect to the white noise. The justification for this notation will become clear in the next paragraph. The random contribution to the evolution \eqref{eq:GeneralOUTkFourierModes} is provided by the stochastic vector term $\widehat{\rd W}_p(t,\bk)$ characterized by the covariance \eqref{eq:CovFourierModesWSpaceTime}. Let us mention a very recent work by \cite{LetMor25} devoted to an equivalent formulation of \cite{ChaGaw03} for a decomposition of the velocity field in a wavelet dyadic basis.

{The unique statistically stationary solution of the stochastic dynamics proposed in \eqref{eq:GeneralOUTkFourierModes} reads
\begin{equation}\label{eq:SolutionStatStatOUTkFourierModes}
\widehat{u}^{(1),\nu}_i(t,\bk_\bn) = \sqrt{\frac{E_\nu^{\text{\tiny{E}}}(k_\bn)}{T_{\bk_\bn}}}\widehat{P}_{ip}(\bk_\bn)\int_{s=-\infty}^t e^{-\frac{t-s}{T_{\bk_\bn}}}\widehat{\rd W}_p(s,\bk_\bn).
\end{equation}
}

From the former expression \eqref{eq:SolutionStatStatOUTkFourierModes}, it is then straightforward to derive the covariance structure of the velocity Fourier modes that are Hermitian symmetric, and characterized by their covariance
\begin{align}\label{eq:CovFourierModesOU}
\E \left[ \widehat{u}_i^{(1),\nu}(t,\bk_\bn)\widehat{u}_j^{(1),\nu}(t+\tau,\bk_\bm)\right] = \frac{L^3_{\text{\tiny{tot}}}}{2}e^{-|\tau|/T_{\bk_\bn}} E_\nu^{\text{\tiny{E}}}(k_\bn)\widehat{P}_{ij}(\bk_\bn)\delta_{\bn,-\bm}^{(3)},
\end{align}
showing that, when compared to the general model provided in \eqref{eq:CovFourierModesPerio}, the temporal structure is given by the function 
\begin{align}\label{eq:FTempCorrOU}
F^{(1)}(\tau)=e^{-|\tau|},
\end{align}
where once again the superscript $^{(1)}$ entering in the particular notation $F^{(1)}$ will become clear next paragraph.

Let us now explore the consequences of the particular temporal kernel $F^{(1)}$ given in \eqref{eq:FTempCorrOU} on the regularity in time of the corresponding velocity field, and more precisely on the scaling behavior of the respective temporal structure function $S_2^{(1),\text{\tiny{T}}}(\tau)$, i.e. the variance of the velocity time increment $\delta_\tau \bu^{(1),\nu}$ \eqref{eq:TimeIncrement}.

Recall that we found the expression \eqref{eq:TemporalS2}, that reads, using the generalized forms of $ E^{\text{\tiny{E}}}(|\bk|)$ \eqref{eq:ModelPSDAnyHNuTo0} and $T_\bk$, \eqref{eq:DefTkBeta} 
\begin{align}\label{eq:TemporalS2AnyH}
S_2^{(1),\text{\tiny{T}}}(\tau)&=\lim_{L_{\text{\tiny{tot}}}\to\infty}\lim_{\nu\to 0}\E\left[ \left|\delta_\tau \bu^{(1),\nu}\right|^2\right]\notag\\
&= \frac{1}{\pi}(1+2H)(3+2H) D_2 \int_{\bk \in \R^3} \left[1-e^{-D_3|\bk|_{1/L}^{2\beta}|\tau|} \right]|\bk|^2 |\bk|_{1/L}^{-(2H+5)} \ \rd \bk .
\end{align}
Similarly to the characteristic time scale $\tau_{\text{c}}(\ell )$ \eqref{eq:RelatTauCTkLtotInfinite}, the temporal structure function $S_2^{(1),\text{\tiny{T}}}(\tau)$ \eqref{eq:TemporalS2AnyH} undergoes a transition at small scales depending on the values of $H$ and $\beta$. We have, while rescaling the dummy integration variable $\bk$ by the appropriate power of the time scale $\tau$ for $H<\beta$:
{\begin{align}\label{eq:TemporalS2AnyHEquiv1}
S_2^{(1),\text{\tiny{T}}}(\tau)&\build{\sim}_{|\tau|\to 0}^{}4(1+2H)(3+2H) D_2 |D_3\tau|^{H/\beta}\int_{\rho=0}^{\infty} \left[1-e^{-\rho^{2\beta}} \right]\rho^{-(2H+1)} \rd \rho.
\end{align}
}
Note that when $H=1/3$ and $\beta=1/2$, one retrieves   \eqref{eq:TemporalS2LimTau0} from the general formulation \eqref{eq:TemporalS2AnyHEquiv1}. For $H>\beta$, one has
\begin{align}\label{eq:TemporalS2AnyHEquiv2}
S_2^{(1),\text{\tiny{T}}}(\tau)\build{\sim}_{|\tau|\to 0}^{}4(1+2H)(3+2H) D_2 |D_3\tau|\int_{\rho=0}^{\infty} \rho^4\sqrt{\rho^2+1/L^2}^{2\beta-(2H+5)} \rd \rho.
\end{align}
The scaling behaviors derived in \eqref{eq:TemporalS2AnyHEquiv1} and \eqref{eq:TemporalS2AnyHEquiv2} pertain to the inertial range. We would also like to explore the behavior of the structure function in the dissipative range, first by taking the limit of vanishing scales $\tau\to 0$, and secondly the limit of vanishing viscosity $\nu\to 0$. When the viscosity $\nu$ is not  zero, instead of \eqref{eq:TemporalS2AnyH}, we have
\begin{align}\label{eq:TemporalS2AnyHVisc}
S_{2,\nu}^{(1),\text{\tiny{T}}}(\tau)&=\lim_{L_{\text{\tiny{tot}}}\to\infty}\E\left[ \left|\delta_\tau \bu^{(1),\nu}\right|^2\right]= 2 \int_{\bk \in \R^3} \left[1-e^{-D_3|\bk|_{1/L}^{2\beta}|\tau|} \right] E_\nu^{\text{\tiny{E}}}(\bk) \ \rd \bk ,
\end{align}
which will clearly behave proportionally to $|\tau|$ for any couple of values of $H$ and $\beta$ due to the behavior of $1-F^{(1)}(\tau/T_\bk)=|\tau|/T_\bk + o(\tau)$ in the vicinity of $\tau=0$. In this regard, the spatio-temporal velocity field $\bu^\nu(t,\bx)$ is not smooth in time and thus does not provide the expected temporal dissipative range scaling in $\tau^2$. This is related to the non-differentiability of the temporal Fourier mode correlation function at zero. Thus, we have to improve the model while proposing another temporal kernel in addition to the simple Markovian model of \cite{ChaGaw03} which has exponential temporal correlation  \eqref{eq:FTempCorrOU}.

Let us close this paragraph by exploring the corresponding temporal spectrum $E_\nu^{(1),\text{\tiny{T}}}(\omega)$. As already considered in \eqref{eq:DefTimeSpectrum}, of the resulting velocity field $\bu^{(1),\nu}$, one has
\begin{align}\label{eq:TimeSpectrumF1}
E_\nu^{(1),\text{\tiny{T}}}(\omega) &= \lim_{L_{\text{\tiny{tot}}}\to\infty}\int_{\tau\in\R}e^{-2i\pi \omega \tau} \E \left[ \bu^{(1),\nu}(t,\bx)\cdot \bu^{(1),\nu}(t+\tau,\bx)\right]\rd \tau \\
&= \int_{\bk \in \R^3} T_\bk\widehat{F}^{(1)}(\omega T_\bk) E_\nu^{\text{\tiny{E}}}(\bk) \ \rd \bk ,\notag
\end{align}
where the one-dimensional Fourier transform $\widehat{F}^{(1)}$ of the temporal kernel $F^{(1)}$ \eqref{eq:FTempCorrOU} is given by
\begin{align}\label{eq:LorentzianF1}
\widehat{F}^{(1)}(\omega)= \frac{2}{1+4\pi^2\omega^2}.
\end{align}
Similarly to the structure function $S_2^{(1),\text{\tiny{T}}}(\tau)$ \eqref{eq:TemporalS2AnyH}, the temporal spectrum $E_\nu^{(1),\text{\tiny{T}}}(\omega)$ \eqref{eq:TimeSpectrumF1} also undergoes a transition depending on the values of $H$ and $\beta$. For $H<\beta$, rescaling the dummy wavenumber $k$ by the appropriate power of the frequency $\omega$, and looking for the equivalent at high frequencies $\omega\to\infty$, we obtain for $H<\beta$
\begin{align}\label{eq:TimeSpectrumF1OmegaInfty}
E^{(1),\text{\tiny{T}}}&(\omega) \equiv \lim_{\nu\to 0} E_\nu^{(1),\text{\tiny{T}}}(\omega)= \int_{\bk \in \R^3} T_{\omega^{\frac{1}{2\beta}} \bk}\widehat{F}^{(1)}(\omega T_{\omega^{\frac{1}{2\beta}} \bk}) E^{\text{\tiny{E}}}(\omega^{\frac{1}{2\beta}} |\bk|) \omega^{\frac{3}{2\beta}} \ \rd \bk ,\notag\\
&\build{\sim}_{\omega\to \infty}^{}2(1+2H)(3+2H) D_2D_3^{H/\beta}\omega^{-\left(1+H/\beta\right)}\int_{0}^\infty \rho^{-\left(2\beta+2H+1 \right)}\widehat{F}^{(1)}(\rho^{-2\beta})\rd \rho \text{ for }H<\beta,
\end{align}
which coincides with \eqref{eq:TimeSpectrumAsympt} when $H=1/3$ and $\beta=1/2$. For $H>\beta$ however, we obtain 
\begin{align}\label{eq:TimeSpectrumF1OmegaInftyTrivial}
E^{(1),\text{\tiny{T}}}(\omega) &\build{\sim}_{\omega\to \infty}^{}\frac{1}{4\pi^3} (1+2H)(3+2H) D_2D_3\omega^{-2}\int_{\bk\in\R^3} |\bk|^2|\bk|_{1/L}^{2\beta-(2H+5)}\rd \bk \text{ for }H>\beta.
\end{align}
Both these asymptotic behaviors \eqref{eq:TimeSpectrumF1OmegaInfty} and \eqref{eq:TimeSpectrumF1OmegaInftyTrivial}  are consistent with the non-differentiability in time of the velocity field. In the first case \eqref{eq:TimeSpectrumF1OmegaInfty}, the regularity in time is determined by the value of $H/\beta$, whereas in the second scenario \eqref{eq:TimeSpectrumF1OmegaInftyTrivial}, this temporal regularity coincides with the one of a Brownian motion, independently of the parameters  $H$ and $\beta$.

\subsubsection{Generalization to a {Markovian} and differentiable-in-time framework}\label{Sec:GeneNLayers}

In the Section above, we have described the statistical properties in time of the Markovian model of \cite{ChaGaw03} which can be seen as an Ornstein-Uhlenbeck type of evolution for Fourier modes \eqref{eq:GeneralOUTkFourierModes}. Fourier modes of the velocity field $\widehat{u}^{(1),\nu}_i(t,\bk)$ at a given wave number $\bk$ are not differentiable in time because they solve a first-order evolution equation driven by a noise which is white-in-time. As a consequence, the temporal paths of $\widehat{u}^{(1),\nu}_i(t,\bk)$ share the same regularity as the Brownian motion and are only H\"{o}lder continuous. This non-differentiable nature of Fourier modes has a counterpart in the physical space: the corresponding velocity field $u^{(1),\nu}_i(t,\bx)$ is not differentiable in time. This is consistent with the behavior of the temporal second-order structure functions $S_{2,\nu}^{(1),\text{\tiny{T}}}(\tau)$, which behaves proportionally to the time scale $\tau$ \eqref{eq:TemporalS2AnyHVisc}, even for a finite viscosity $\nu>0$. If the velocity field were differentiable, we would expect instead a quadratic behavior, i.e. $S_{2,\nu}^{(1),\text{\tiny{T}}}(\tau)=O_\nu(\tau^2)$ where $O_\nu(\tau^2)$ stands for a term of order $\tau^2$ with a {viscosity-dependent} multiplicative factor. The necessity of proposing a dynamic capable of achieving this goal in a {Markovian} way is of tremendous importance, {both from a physical and numerical perspectives}.  To do so, we will exploit an idea first proposed by \cite{Saw91} and then generalized to an infinitely differentiable framework by \cite{VigFri20} that consists of replacing the white noise entering in the dynamics \eqref{eq:GeneralOUTkFourierModes} by a continuous random force, which is itself of Ornstein-Uhlenbeck type and/or itself governed by a dynamics stirred by a continuous random force, which would ensure the finiteness of the first and/or subsequent time derivatives.

We are now in position to introduce the evolution for the Fourier mode $\widehat{u}^{(N),\nu}_i(t,\bk)$.%, in place of \eqref{eq:GeneralOUTkFourierModes}, reads as follows. 
 Given an integer $N\ge 2$, we consider the following set of $N$ embedded evolution equations:
\begin{align}
 \label{eq:OU_embedded_N_1}
 \frac{\rd \widehat{u}^{(N),\nu}_i(t,\bk)}{\rd t} &= -\frac{\sqrt{4N}}{T_\bk}\widehat{u}^{(N),\nu}_i(t,\bk)+\widehat{f}^{(N-1)}_i(t,\bk)\;\\
 %\label{eq:OU_embedded_two_2}
 \frac{\rd \widehat{f}^{(N-1)}_i(t,\bk)}{\rd t} &= -\frac{\sqrt{4N}}{T_\bk}\widehat{f}^{(N-1)}_i(t,\bk)+\widehat{f}^{(N-2)}_i(t,\bk)\;\\
 &\vdots\notag\\ 
 \frac{\rd\widehat{f}^{(2)}_i(t,\bk)}{\rd t} &= -\frac{\sqrt{4N}}{T_\bk}\widehat{f}^{(2)}_i(t,\bk)+ \widehat{f}^{(1)}_i(t,\bk)\;\\
 \rd\widehat{f}^{(1)}_i(t,\bk)&= -\frac{\sqrt{4N}}{T_\bk}\widehat{f}^{(1)}_i(t,\bk) \rd t + \sqrt{q^{(N)}E_\nu^{\text{\tiny{E}}}(\bk)}\widehat{P}_{ip}(\bk)\widehat{\rd W}_p(t,\bk)\;,
\label{eq:OU_embedded_N_N}
\end{align}
where we introduce in the right-hand side of  \eqref{eq:OU_embedded_N_N} the parameter $q^{(N)}$ defined by
\begin{align} \label{eq:ValueqN}
q^{(N)}&=\frac{1}{2}\frac{1}{ \int_{\mathbb R}\frac{1}{\left[1+\pi^2\frac{T_\bk^{2}}{N}\omega^2\right]^N} d\omega}\left(\frac{4N}{T_\bk^2}\right)^N =\frac{1}{2}\frac{T_\bk\sqrt{\pi} \Gamma(N)}{\sqrt{N} \Gamma(N-1/2)}\left(\frac{4N}{T_\bk^2}\right)^N.
\end{align}

{Let us now clarify the Markovian nature of the evolution proposed in the set of equations \eqref{eq:OU_embedded_N_1} to \eqref{eq:OU_embedded_N_N}. Although the very evolution of the respective Fourier mode $\widehat{u}^{(N),\nu}_i(t,\bk)$ cannot be considered as Markovian in itself, the $N$-dimensional vector made up of that Fourier mode and the $(N-1)$ remaining underlying layers $\widehat{f}^{(m)}_i(t,\bk)$ for $N-1\ge m\ge 1$, as it is defined in  \eqref{eq:DefFTotalMarkov} considered in the numerical Section \ref{Sec:NumericalSchemes}, can be formulated in a Markovian way. The respective matrix evolution is given in \eqref{eq:MarkEvolFullF}.}

We show in Appendix \ref{App:CovEmbeddedModel} that, in the statistically stationary regime, one has
\begin{align} \label{eq:CovHatUFiniteN}
\lim_{t\to\infty} \E\left[\widehat{u}^{(N),\nu}_i(t,\bk_\bn)\widehat{u}^{(N),\nu}_j(t+\tau,\bk_\bm)\right]= \frac{L^3_{\text{\tiny{tot}}}}{2}F^{(N)}(\tau/T_{\bk_\bn})E_\nu^{\text{\tiny{E}}}(k_\bn)\widehat{P}_{ij}(\bk_\bn)\delta^{(3)}_{\bn,-\bm},
\end{align}
where the temporal correlation function $F^{(N)}$ is defined by
 \begin{align}\label{eq:FNTAU}
 F^{(N)}(\tau) = \frac{2 |\sqrt{N}\tau|^{N-1/2}
 K_{N-1/2}\left( 2|\sqrt{N}\tau|\right)}{\Gamma(N-1/2)}.
 \end{align}
In \eqref{eq:FNTAU}, $K_n(x)$ is the modified Bessel function of the second kind. We recall that the expression of $F^{(N)}$ \eqref{eq:FNTAU} should be considered for $N\ge 2$. In particular, considering $F^{(N)}$ \eqref{eq:FNTAU} for $N=1$ does not coincide with the expression $F^{(1)}$ given in \eqref{eq:FTempCorrOU}. In this situation, the Fourier mode $\widehat{u}^{(N),\nu}_i(t,\bk_\bn)$ is differentiable $(N-1)$-times, and we provide the expression of the variance of the $p$-derivatives \eqref{eq:ExprePDerivativesUN}, which is finite as long as $p\le N-1$.

{Surprisingly, the expression of the correlation function $F^{(N)}$ \eqref{eq:FNTAU} coincides exactly with the one of a Mat\'ern Gaussian process, which have been extensively used in machine learning theory, as reviewed by \cite{WilRas06}. More recently, such Gaussian processes have been exploited in a turbulent modeling context by \cite{LilSyk17}. To our understanding, this type of covariance has been called in these formerly cited investigations for the necessity of considering differentiable Gaussian processes. Our approach furthermore shows that this covariance function of Mat\'ern type  \eqref{eq:FNTAU} can be realized in a Markovian fashion, following the set of embedded Ornstein-Uhlenbeck evolutions provided in equations \eqref{eq:OU_embedded_N_1} to  \eqref{eq:OU_embedded_N_N}. In this sense, it can be viewed as a new way to interpret the nature of Mat\'ern Gaussian processes.}

Moreover, in the limit of an infinite number of embedded layers $N\to\infty$, one has
\begin{align}\label{eq:FInfiniteNTAU}
 F^{(\infty)}(\tau)= \lim_{N\to\infty}F^{(N)}(\tau) = e^{-\tau^2}.
 \end{align}
As a consequence, one obtains
\begin{align} \label{eq:ConvTowardGaussianKernel}
\lim_{N\to\infty}\lim_{t\to\infty} \E\left[\widehat{u}^{(N),\nu}_i(t,\bk_\bn)\widehat{u}^{(N),\nu}_j(t+\tau,\bk_\bm)\right]=\frac{L^3_{\text{\tiny{tot}}}}{2}e^{-(\tau/T_{\bk_\bn})^2}E_\nu^{\text{\tiny{E}}}(k_\bn)\widehat{P}_{ij}(\bk_\bn)\delta^{(3)}_{\bn,-\bm}.
\end{align}
 
{Let us now explore the consequences of the particular temporal kernel $F^{(N)}$ given in \eqref{eq:FNTAU} (resp. $F^{(\infty)}$ given in \eqref{eq:FInfiniteNTAU}) on the regularity in time of the corresponding velocity field $u^{(N),\nu}_i(t,\bx)$ (resp. ${u^{(\infty),\nu}_i(t,\bx)}$). We gather in Appendix \eqref{App:S2andEMultiLayered} the detailed derivation necessary to justify them.}

{Let us first focus on the temporal structure functions, i.e. the variance of the velocity time increment \eqref{eq:TimeIncrement}. Using the generalized forms of $ E^{\text{\tiny{E}}}(k)$ \eqref{eq:ModelPSDAnyHNuTo0} and $T_k$, \eqref{eq:DefTkBeta}, the expression \eqref{eq:TemporalS2} reads 
\begin{align}\label{eq:DefTemporalS2AnyHFNAndInf}
S_2^{(N),\text{\tiny{temp}}}(\tau)&=\lim_{L_{\text{\tiny{tot}}}\to\infty}\lim_{\nu\to 0}\E\left[ \left|\delta_\tau \bu^{(N),\nu}\right|^2\right],
\end{align}
which includes the limiting case $N=\infty$.
}

{Similarly to the temporal structure function \eqref{eq:TemporalS2AnyH} associated with the exponential kernel \eqref{eq:FTempCorrOU}, $S_2^{(N),\text{\tiny{temp}}}(\tau)$  \eqref{eq:DefTemporalS2AnyHFNAndInf} eventually undergoes a transition at small scales, but for a different range of values of $H$ and $\beta$. We indeed obtain for any $N$, including $N=\infty$, 
\begin{align}\label{eq:SimplifiedTemporalS2AnyHEquiv1FN}
S_2^{(N),\text{\tiny{temp}}}&(\tau)\build{\propto}_{|\tau|\to 0}^{}|D_3\tau|^{H/\beta} \mbox{ for }H<2\beta.
\end{align}
We provide in \eqref{eq:TemporalS2AnyHEquiv1FN} and \eqref{eq:TemporalS2AnyHEquiv1FNInfinite} an explicit expression of the multiplicative constant. Note that, for a given $\beta$, the range of accessible spatial regularity $H$ is wider (i.e. $H<2\beta$) in \eqref{eq:SimplifiedTemporalS2AnyHEquiv1FN} than what was found (i.e. $H<\beta$) in \eqref{eq:TemporalS2AnyHEquiv1}. The behavior of the structure function for $H>2\beta$ is provided in \eqref{eq:TemporalS2AnyHFNTrivial}.}

{Importantly, seeking for the limiting behavior at small scales $\tau\to 0$ without taking first the vanishing viscosity limit, we now obtain as soon as $N\ge 2$ 
\begin{align}\label{eq:SimplifiedTemporalS2AnyHViscFN}
S_{2,\nu}^{(N),\text{\tiny{temp}}}(\tau)\build{\propto}_{|\tau|\to 0}^{} \left| D_3\tau\right|^2.
\end{align}
Once again, a more precise expression than \eqref{eq:SimplifiedTemporalS2AnyHViscFN} is given in \eqref{eq:TemporalS2AnyHViscFN}. Thus, it now behaves proportionally to $\tau^2$ for any pair of values of $H$ and $\beta$: the obtained velocity field $\bu^{(N),\nu}(t,\bx)$ is now smooth in time as long as $\nu>0$ and $N\ge 2$.}

{Considering now the associated temporal spectrum $E_\nu^{(N),\text{\tiny{T}}}(\omega)$ defined by
\begin{align}\label{eq:DefTimeSpectrumFN}
E_\nu^{(N),\text{\tiny{T}}}(\omega) &= \lim_{L_{\text{\tiny{tot}}}\to\infty}\int_{\tau\in\R}e^{-2i\pi \omega \tau} \E \left[ \bu^{(N),\nu}(t,\bx)\cdot \bu^{(N),\nu}(t+\tau,\bx)\right]\rd \tau, 
\end{align}
we obtain for $H<(2N-1)\beta$
\begin{align}\label{eq:SimplifiedTimeSpectrumFNOmegaInfty}
E^{(N),\text{\tiny{T}}}(\omega) &\equiv \lim_{\nu\to 0} E_\nu^{(N),\text{\tiny{T}}}(\omega)\build{\propto}_{\omega\to \infty}^{}\omega^{-\left(1+H/\beta\right)},
\end{align}
with a more precise expression given in \eqref{eq:TimeSpectrumFNOmegaInfty}, and in \eqref{eq:TimeSpectrumFNInfinityOmegaInfty} for the limiting case $N\to \infty$, The case $H>(2N-1)\beta$ is treated in \eqref{eq:TimeSpectrumFNOmegaInftyTrivial}.}

\subsection{A second comparison concerning the temporal structure of DNS and the Gaussian model}\label{Sec:CompTempStatDNSModel}

In Section \ref{Sec:CompSpatStatDNSModel}, we compared have the spatial structure of the model velocity field $\bu^{\nu}(t,\bx)$ with the spatial structure of the DNS velocity field $\bu^{\text{\tiny{DNS}},\nu}(t,\bx)$. In this section, we would like to compare the temporal structure of these velocity fields.  Once again, the DNS data are obtained from the Johns Hopkins turbulence database, where 5028 time steps are stored among the evolution in a statistically stationary regime. We have downloaded the three velocity components of $\bu^{\text{\tiny{DNS}},\nu}(t,\bx)$ in a limited region of space defined by three planes of coordinates $x=0$, $y=0$ and $z=0$. 

\begin{figure}
\centering
\includegraphics[scale = 0.7]{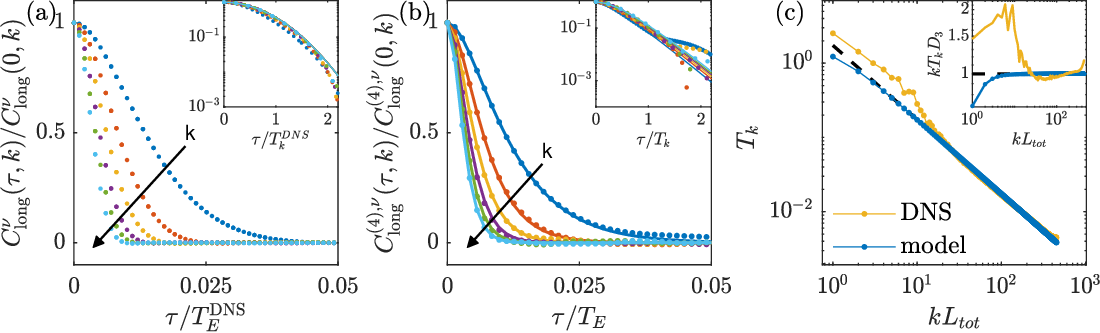}
\caption{Temporal correlation structure of the longitudinal velocity Fourier modes \eqref{eq:Def1dLongFourierModes}. (a) Longitudinal correlation function $C_{\text{\tiny{long}}}^{\nu}(\tau,k_n)$ \eqref{eq:DefTimeCorrLongFM} of the DNS velocity field $\widehat{u}_{\text{\tiny{long}}}^{\text{\tiny{DNS}},\nu}(t,k_n)$, for $k_n L_{\text{\tiny{tot}}}=42,\ 75,\ 107,\ 141,\ 173,\ 205$. The characteristic large time scale $T_E^{\text{\tiny{DNS}}} = 1.99 $ is defined in the \href{https://turbulence.idies.jhu.edu/docs/isotropic/README-isotropic.pdf}{readme file} of the DNS. Inset: Re-scaled correlation functions {in semi-log scale} for the same wave numbers $k_n$, with $D_3$ provided in the text {and the theoretical gaussian expression in solid line} by \cite{GorBal21}. (b) Similar representation of the longitudinal correlation function {$C_{\text{\tiny{long}}}^{(4),\nu}(\tau,k_n)$} \eqref{eq:DefTimeCorrLongFM} as in (a) but for the Gaussian model {$\widehat{u}_{\text{\tiny{long}}}^{(4),\nu}(t,k_n)$}. Solid lines represent theoretical predictions while dot-lines are numerical estimations for the same wavenumbers $k_n$ as (a). The characteristic large time scale $T_E = 1.43 $ is defined by $T_E :=  \sqrt{3}\frac{L_{\text{\tiny{int}}}}{\sigma^\nu}$ with $L_{\text{\tiny{int}}}$  being the integral time scale defined in \eqref{eq:RelatLtoLint} Inset: Re-scaled correlation functions {in semi-log scale} for the same wave numbers $k_n$, with the same $D_3$ as for DNS and provided in the text. (c) Numerical estimation of the characteristic time scale $T_{k_n}$ of the time correlation functions displayed in (a) and (b). We superimpose the numerical estimation of $T_{k_n}$ with a dashed black line given by the large $k$ asymptotic $(D_3 k)^{-1}$ of  the theoretical expression \eqref{eq:DefTkBeta}, with relevant free parameters provided in the text. Inset: re-scaled estimated $T_{k_n}$ by $(D_3 k)^{-1}$, the asymptotical behavior of $T_k$ at large $k$.  }
\label{Fig:TempCorr3D}
\end{figure}

We begin by estimating on the DNS field the time correlation of Fourier modes $\widehat{\bu}^{\text{\tiny{DNS}},\nu}(t,\bk_\bn)$ for several values of the wave vector $\bk_\bn$, as it has been already done by \cite{GorBal21}. To be more precise, and because we have downloaded data only in a limited region of space, we will be focusing on the correlation in time of the one-dimensional longitudinal Fourier modes. We derive in Appendix \ref{App:Rel3d1dTimeCorrelation} the precise relationship between the time-correlation structure on the one-dimensional longitudinal Fourier modes 
\begin{align}\label{eq:DefTimeCorrLongFM}
C_{\text{\tiny{long}}}^{(N),\nu}(\tau,k_n) = \frac{\E \left[ \widehat{u}_{\text{\tiny{long}}}^{(N),\nu}(t,k_n)\widehat{u}_{\text{\tiny{long}}}^{(N),\nu}(t+\tau,-k_n)\right]}{\E \left[ \left|\widehat{u}_{\text{\tiny{long}}}^{(N),\nu}(t,k_n)\right|^2\right]} \equiv F^{(N)}_{\text{\tiny{long}},k_n}\left(\tau\right),
\end{align}
and its corresponding three-dimensional counterpart that is fully characterized by the temporal kernel $F$ entering in \eqref{eq:CovFourierModesPerio}. By virtue of statistical homogeneity and isotropy, the one-dimensional longitudinal Fourier mode $\widehat{u}_{\text{\tiny{long}}}^{(N),\nu}(t,k_n)$ can be defined as
\begin{align}\label{eq:Def1dLongFourierModes}
\widehat{u}_{\text{\tiny{long}}}^{(N),\nu}(t,k_n)= \int_{[-L_{\text{\tiny{tot}}}/2\ ;\ L_{\text{\tiny{tot}}}/2]}e^{-2i\pi k_n x}u^{(N),\nu}_1(t,x,0,0)\rd x. 
\end{align}
The solid lines of Fig. \ref{Fig:TempCorr3D}(b) are the theoretical predictions of $F^{(N)}_{\text{\tiny{long}},k_n}(\tau)$ for the above mentioned $k_n$, computed in Appendix \ref{App:Rel3d1dTimeCorrelation} 
\begin{align}\label{eq:RelFlongF3D}
F_{\text{\tiny{long}}, k_n}^{(N)}\left(\tau\right) = L_{\text{\tiny{tot}}} \int\limits_0^{+\infty} \frac{\rho^2}{k_n^2+\rho^2}\frac{E^{\text{\tiny{E}}}_\nu\lp\sqrt{k_n^2+\rho^2}\rp}{2} F^{(N)}\lp \frac{\tau}{T_{\sqrt{k_n^2 + \rho^2}}}\rp\ 2\pi\rho\ \rd \rho \ .
\end{align}
In the case of $F^{(\infty)}$ and for $\beta = 1/2$, it can be written as
\begin{equation}
F^{(\infty)}_{\text{\tiny{long}},k_n} = L_{\text{\tiny{tot}}} e^{-\tau^2 D_3^2 \lp k_n^2 + L^{-2}\rp } \int\limits_0^{+\infty} \frac{\rho^2}{\rho^2+k_n^2} \frac{E_\nu^{\text{\tiny{E}}}\lp\sqrt{\rho^2+k_n^2}\rp}{2}e^{-D_3^2 \tau^2 \rho^2} \ 2\pi\rho \ \rd \rho \ .
\end{equation}
To estimate $D_3$ from the DNS, we therefore assume the integral to vary slowly with $k_n$ and fit $F^{(\infty)}_{\text{\tiny{long}},k_n} $ with a Gaussian function. This assumption may be verified by noting that all curves in \ref{Fig:TempCorr3D}(b) collapse while rescaling by $T_k$. Following this method, the estimated value of $D_3$ is then $D_3 = 3.62$.

We show in Fig. \ref{Fig:TempCorr3D}(a) the results of our numerical estimation of the time correlation function {of the DNS, $C_{\text{\tiny{long}}}^{\nu}(\tau,k_n)$ }\eqref{eq:DefTimeCorrLongFM} for several wave-number amplitudes $k_n$ and as a function of the time-lag $\tau$. At vanishing time lag $\tau=0$, all curves coincide with unity by construction, and then decrease toward 0 (they decorrelate) as $\tau$ increases. The characteristic time scale {$T_{k_\bn}^{\text{\tiny{DNS}}}$} of decorrelation is thus clearly a decreasing function of the wave number $k_n$. As has been observed by \cite{GorBal21}, the very shape of the correlation function is close to a Gaussian function, and so, to estimate the characteristic time scale {$T_{k_n}^{\text{\tiny{DNS}}}$}, we perform a similar Gaussian fit as by \cite{GorBal21}. In a straightforward manner, for each accessible $1/L_{\text{\tiny{tot}}}<k_n<512/L_{\text{\tiny{tot}}}$, we find the best possible {$T_{k_n}^{\text{\tiny{DNS}}}$} such that the temporal correlation function {$C_{\text{\tiny{long}}}^{\nu}(\tau,k_n)$} \eqref{eq:DefTimeCorrLongFM} resembles the explicit Gaussian function {$\exp(-(\tau/T_{k_n}^{\text{\tiny{DNS}}})^2)$} and reproduce the result of this fitting procedure in Fig. \ref{Fig:TempCorr3D}(c) using an orange line. We can indeed observe that {$T_{k_n}^{\text{\tiny{DNS}}}$} is a decreasing function of the wave number $k_n$, and follows a power-law, as has clearly been evidenced by the linear trends in this doubly logarithmic representation. We can also identify two ranges, below and above $k_nL_{\text{\tiny{tot}}}\approx 10$. This transition has already been observed in several studies \citep{KanIsh99,FavGod10,GorBal21} and remains unexplained, as far as we know. We believe that at large scales, i.e. at low wave numbers, the flow is sensitive to boundary conditions and thus cannot be considered universal. At higher wave numbers $k_n L_{\text{\tiny{tot}}}$ greater than 10, the power decrease is consistent with the expected behavior given in \eqref{eq:DefTkBeta} with $\beta=1/2$ (governing the power-law exponent) and $D_3 = 3.62$. We superimpose with a dashed black line the proposed function dependence of $T_{k_n}$ \eqref{eq:DefTkBeta} {in the limit of large $k_n$}, with these aforementioned free parameters and clearly reproduce the decreasing behavior. To check whether such a parameterization of the characteristic time scale reproduces the behaviors of Fourier modes time correlation functions, we add in the inset of Fig. \ref{Fig:TempCorr3D}(a) a representation of {$C_{\text{\tiny{long}}}^{\nu}(\tau,k_n)$} \eqref{eq:DefTimeCorrLongFM} as a function of the re-scaled time variable  {$T_{k_n}^{\text{\tiny{DNS}}}$ in semi-log scale}, and observe that {all correlation functions collapse in a satisfactory manner on the expected theoretical curve}.

For comparison against our modeling approach, we consider the Gaussian velocity field $u^{{(4)},\nu}_i(t,\bx)$, i.e. for {$N=4$}. It is fully characterized in the statistically stationary regime by the covariance structure of its Fourier modes {$\widehat{u}^{(4),\nu}_i(t,\bk_\bn)$} which is provided in \eqref{eq:CovHatUFiniteN}. We have considered an evolution based on {four} layers {$N=4$} that ensures a smooth behavior in time for a finite viscosity, as it is shown in \eqref{eq:TemporalS2AnyHViscFN}. This choice is dictated by the necessity of ensuring smooth behavior in time of the velocity field, corresponding to bounded velocity time-derivatives (in this case, {$\partial_t u^{(N),\nu}_i(t,\bx)$ have a finite variance for all integer $N>1$}), as is observed in DNS. As we explain in Chapter \ref{Sec:NumericalSchemes}, each layer needs to be stored in memory, which can become prohibitive given the size of the boxes, which corresponds to three-components of size $1024^3$ collocation points for each layer. Besides considering $H=1/3$, the free parameters $D_2$, $L$ and $\eta_d$ entering in the functional form of the longitudinal PSD \eqref{eq:ModelPSD1dLongAnyH} have the same values as the ones chosen to describe the spatial structure of the velocity field, as they are given in Section \ref{Sec:CompSpatStatDNSModel}. The remaining free parameters $D_3$ and $\beta$ that appear in the characteristic correlation time scale $T_k$ of Fourier modes \eqref{eq:DefTkBeta} are the ones that we have estimated before, corresponding to $\beta=1/2$ and $D_3 = 3.62$. We would like to recall that the value of $D_3$ is not universal because it is related, when $\beta=1/2$, to the standard deviation of the velocity. An efficient and exact in distribution numerical scheme is proposed in Section \ref{Sec:NumericalSchemes} in order to give instances in space and time of the model field {$u^{(N),\nu}_i(t,\bx)$ for all integer $N>1$}. {Overall, the simulation of the vector field over more than five thousands time steps took approximatively two days using 32 cpus.}

As we did for the DNS velocity field, we display in Fig. \ref{Fig:TempCorr3D}(b) the temporal longitudinal correlation functions {$C_{\text{\tiny{long}}}^{(N),\nu}(\tau,k_n)$} \eqref{eq:DefTimeCorrLongFM} for the modeled Gaussian velocity field {$\bu^{(4),\nu}(t,\bx)$}. Although being known to be different from the DNS in a theoretical point of view and provided in \eqref{eq:CovHatUFiniteN} (once recasted into the longitudinal context \eqref{eq:RelFlongF3D}), the estimations of {$C_{\text{\tiny{long}}}^{(4),\nu}(\tau,k_n)$} for various wave numbers $k_n$ displays similar trends as for the DNS. In this case, the collapse is not perfect because we are considering only the longitudinal correlation functions, which necessitate a transformation given in 
\eqref{eq:RelFlongF3D}) when going from the three dimensional formulation to the one dimensional setting.  We display in Fig. \ref{Fig:TempCorr3D}(c) the time $T_k$ as a function of $k$, for the model for which Fourier mode temporal correlation is prescribed using blue symbols, and the measured Fourier mode correlation time for DNS in orange. We observe a very satisfactory collapse of the estimations of $T_k$ from DNS and the model. We also superimpose {in dashed line} the asymptotical limit {$\lp D_3 k_n\rp^{-1}$} using the form provided {in \eqref{eq:DefTkBeta}. In} the inset, we display the re-scaled correlation functions by the asymptotic behavior of the imposed characteristic time scale { $\lp D_3 k_n\rp^{-1}$ }\eqref{eq:DefTkBeta} at large $k_n$.

 \begin{figure}
\centering
\includegraphics[scale = 0.7]{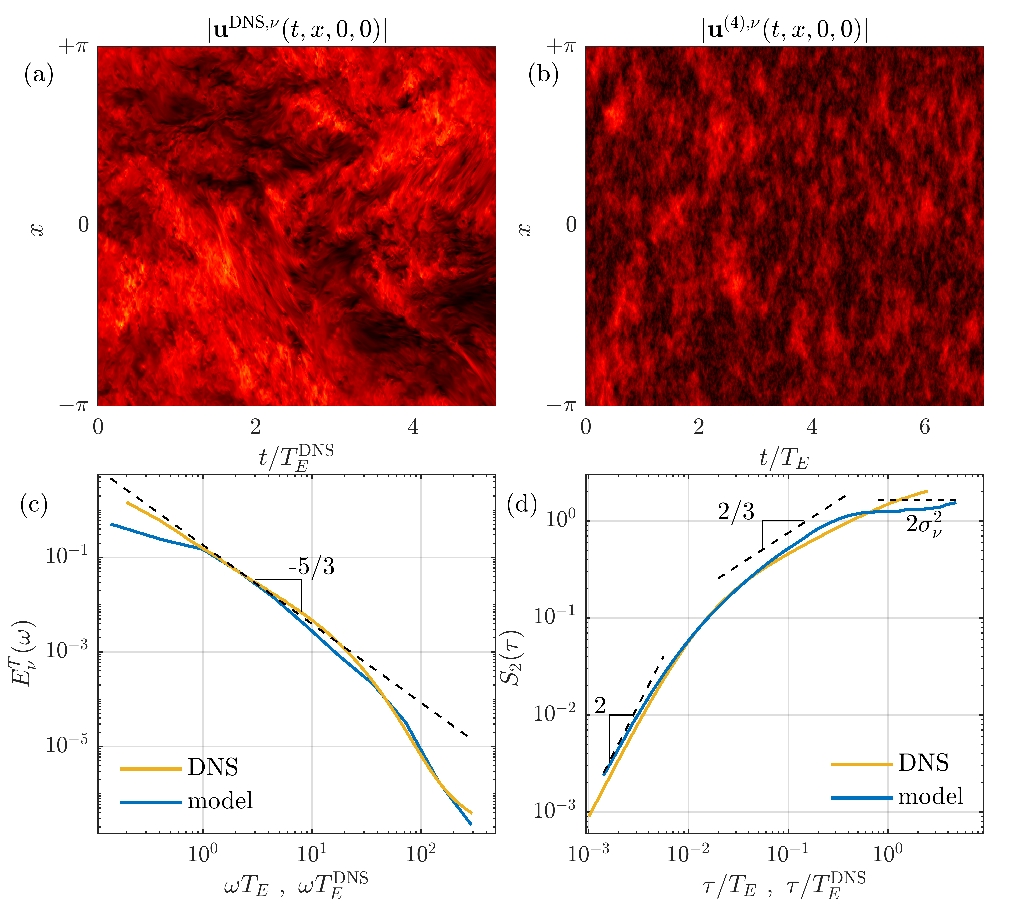}
\caption{Direct comparison of the temporal structure of the DNS and modeled velocity fields, and estimation of PSDs and second-order structure functions. (a) Spatio-temporal representation of the DNS velocity field $|\bu^{\text{\tiny{DNS}},\nu}(t,\bx)|$ along the spatial line $\bx\in([-\pi,\pi],0,0)$ and across time $t\in[0,5027\Delta t]$ where the time stepping $\Delta t$ is provided by the Hopkins database. We use the same characteristic large time scale $T_E^{\text{\tiny{DNS}}}$ as in Fig. \ref{Fig:TempCorr3D} to adimensionalize time. (b) Similar representation as in panel (a) but for the model {$|\bu^{(4),\nu}(t,\bx)|$} with a characteristic large time scale $T_E$ { of the order of $T_E^{\text{\tiny{DNS}}}$}. We use the same colorbar for panels (a) and (b). (c) Estimation of the temporal power spectral density $E_\nu^{\text{\tiny{T}}}(\omega)$ \eqref{eq:DefTimeSpectrum} obtained as the variance of the temporal Fourier modes (see text). We use orange for DNS and blue for the model. The dashed black line corresponds to the power-law given in \eqref{eq:TimeSpectrumFNOmegaInfty} without any fitting parameters. (d) Estimation of the second-order temporal structure function $S_2^{\text{\tiny{T}},\nu}(\tau)$ (given by \eqref{eq:TemporalS2} in the inviscid limit), same colors as in (c). Dashed black lines corresponds to two times the variance at large time lags, to \eqref{eq:TemporalS2AnyHEquiv1FN} in the inertial range, and to \eqref{eq:TemporalS2AnyHViscFN} in the dissipative range.}
\label{Fig:TempSnaps}
\end{figure}

We now present in Fig. \ref{Fig:TempSnaps} instances of the time evolution of both DNS and Gaussian model velocity fields, with the statistical estimation of the respective temporal spectra and structure functions. We begin in Fig. \ref{Fig:TempSnaps}(a) by displaying the time evolution of the velocity norm $|\bu^{\text{\tiny{DNS}},\nu}(t,\bx)|$, along the line $\bx\in([-\pi,\pi],0,0)$ and across time $t\in[0,5028\Delta t]$, where the time stepping $\Delta t$ is provided by the Hopkins database. We compare with a similar representation in \ref{Fig:TempSnaps}(b) of the same quantity but for the model {$|\bu^{(4),\nu}(t,\bx)|$}. The same colorbar has been used in both cases. Although the spatial structures of these two velocity fields are very similar, up to some filamentary structures that are not reproduced by the model (see the discussion in Section \ref{Sec:CompSpatStatDNSModel} and Figure \ref{Fig:SpatialSnaps}), the spatio-temporal representation of the velocity norm exhibits a key difference between DNS and the model: for the DNS, we can observe that large scale structures of typical size the integral length scale are swept by the flow, whereas it is not the case in the model.  Although not clearly evidenced in this representation, we expect the absence of this sweeping property to be true for eddies at any scale. Thus, while the statistical correlation structures of Fourier modes are well reproduced by the model, as evidenced in Fig. \ref{Fig:TempCorr3D}, the model misses one of the important aspects of the sweeping phenomenon, which is the advection of the large scales, as observed in DNS. We invite the reader to the conclusion where this necessary modeling step is proposed as a perspective.

Our statistical estimation of the temporal power spectrum density $E_\nu^{\text{\tiny{T}}}(\omega)$ defined in \eqref{eq:DefTimeSpectrum} is shown in Fig. \ref{Fig:TempSnaps}(c). To do so, we performed a discrete one-dimensional Fourier transform in time of the profile of the scalar product of velocity fields at a fixed position in space, as required in the definition \eqref{eq:DefTimeSpectrum}. The obtained evolution in time is not periodic, and in doing so we introduce a spurious discontinuity in the picture. We have checked that windowing this profile aimed at minimizing the consequences of this discontinuity is not altering the observed behaviors in the inertial range of scales, but is slightly changing the results in the respective dissipative range (data not shown).  As a result, the temporal spectra correspond to the variance of the (temporal) Fourier modes at each accessible frequency. The variances of the (temporal) Fourier modes of the velocity fields $\bu^{\text{\tiny{DNS}},\nu}(t,\bx)$ and {$\bu^{(N),\nu}(t,\bx)$}, obtained of the same statistical sample previously described, as functions of the frequency, are displayed doubly logarithmically in Fig. \ref{Fig:TempSnaps}(c). We can see that with previously given free parameters, the temporal spectra superimpose in a very satisfactory manner in the inertial range, and follow a typical $\omega^{-5/3}$ power-law decrease. 

We finish with the estimation of the second-order temporal structure function $S_2^{\text{\tiny{T}},\nu}(\tau):= \E\left[ \lp \bu^\nu\lp t+\tau,\bx\rp - \bu^\nu\lp t,\bx\rp \rp^2\right]$, as defined in \eqref{eq:TemporalS2} in the  vanishing viscosity limit, and display our results in Fig. \ref{Fig:TempSnaps}(d). The trends are consistent with the observations made for the temporal spectra. Both DNS and model velocity fields exhibit a power-law behavior {both} in the inertial { and the dissipative} ranges, with {respective power-law exponents indicated on the figure}. To see this, we superimpose the prediction made in \eqref{eq:TemporalS2AnyHEquiv1FN} with a black dashed line. Note that in the inertial range of scales, predictions depend only weakly on the value of the number of layers N to define the model.

To conclude this comparison, {let us mention that additional comparisons could be done. In particular, we could also display the full spatio-temporal correlation function, as it has been done by \cite{ClaCob15}, and  the frequency-wave spectrum as estimated by \cite{KatGan15}, for both the DNS and the model. We are keeping this for further investigations. }We would like also to mention that while the statistical structure of the DNS is fairly satisfactory (at least up to the aforementioned second-order statistical characterization), our model does not capture in a fully appropriate way the advection of the eddies by some large scale flow, as it is clear from the inspection of Fig. \ref{Fig:TempSnaps}(a) and Fig. \ref{Fig:TempSnaps}(b). To our knowledge, there is no model capable of doing this except some recent propositions by \cite{ArmVic25}, where the temporal structure of the advecting velocity field, although simplistic in terms of regularity, is defined recursively through scales using an inverse Lagrangian flow map at each scale of the construction. We will return to these possible perspectives in the final Section. 

\section{Numerical method}\label{Sec:NumericalSchemes}

\subsection{Remarks concerning the Discrete Fourier Transform (DFT)}\label{Sec:DFT}

In all simulations, we consider a finite number of Fourier modes, and thus, we need to truncate in an appropriate manner the Fourier series  and perform an associated Discrete Fourier Transform (DFT). For full benefit of the Fast Fourier Transform (FFT) algorithm, we will be considering an even number of modes $N_x$ in each direction, which will furthermore be taken as a power of 2, i.e. $N_x=2^p$, with $p$ being a given integer. In this spirit, over the grid $\bx_\bn=\bn L_{\text{\tiny{tot}}}/N_x$ with  $\llbracket\ -N_x/2 \ ;\ N_x/2-1\ \rrbracket^3$, here and further, we define the DFT as, for $\bk_\bm=\bm/L_{\text{\tiny{tot}}}$ with $\bm\in\llbracket\ -N_x/2 \ ;\ N_x/2-1\ \rrbracket^3$,
\begin{align}\label{eq:DefDFTVelocity}
\widehat{u}^\nu_i(t,\bk_\bm) = \left(\frac{L_{\text{\tiny{tot}}}}{N_x}\right)^3\sum_{ \bn\in\llbracket\ -N_x/2 \ ;\ N_x/2-1\ \rrbracket^3} e^{-2i\pi \bk_\bm\cdot \bx_n} u^\nu_i(t,\bx_n),
\end{align}
with corresponding inverse DFT
\begin{align}\label{eq:DefinverseDFTVelocity}
u^\nu_i(t,\bx_\bn) = \frac{1}{L^3_{\text{\tiny{tot}}}}\sum_{\bm\in\llbracket\ -N_x/2 \ ;\ N_x/2-1\ \rrbracket^3} e^{2i\pi \bk_\bm\cdot \bx_\bn} \widehat{u}^\nu_i(t,\bk_\bm).
\end{align}

\subsection{Numerical schemes}

\subsubsection{Numerical scheme for the Markovian evolution ($N=1$)}

Consider first the \OUp\ given by \eqref{eq:GeneralOUTkFourierModes}, and recall that in the \ssr\ its solution is given by \eqref{eq:SolutionStatStatOUTkFourierModes}. {The numerical scheme we use is derived from the solution \eqref{eq:SolutionStatStatOUTkFourierModes} by computing analytically the time-evolution of the field $\widehat{u}^{(1),\nu}_i(t,\bk)$ at time $t_{m+1} := t_m + \delta t$ as the function of the solution at time $t_m$:} 
\begin{equation}\label{eq:SolutionGeneralOUTkFourierModesGeneral}
\widehat{u}^{(1),\nu}_i(t_{m+1},\bk) = e^{-\frac{\delta t}{T_\bk}}\widehat{u}^{(1),\nu}_i(t_m,\bk) + \sqrt{\frac{E_\nu^{\text{\tiny{E}}}(k)}{T_\bk}}\widehat{P}_{ip}(\bk)\int_{s=t_{m}}^{t_{m+1}} e^{-\frac{t_{m+1}-s}{T_\bk}}\widehat{\rd W}_p(s,\bk).
\end{equation}
The two terms on the right-hand side of \eqref{eq:SolutionGeneralOUTkFourierModesGeneral} are independent Gaussian random variables, therefore the values of $\widehat{u}^{(1),\nu}_i(t_{m},\bk)$ can be computed recursively if one is able to sample the sequence of independent Gaussian random variables $\left(\int_{s=t_{m}}^{t_{m+1}} e^{-\frac{t_{m+1}-s}{T_\bk}}\widehat{\rd W}_p(s,\bk)\right)_{m\ge 0}$. It suffices to compute the expected value 
\[
\E\left[\int_{s=t_{m}}^{t_{m+1}} e^{-\frac{t_{m+1}-s}{T_\bk}}\widehat{\rd W}_p(s,\bk)\right]=0
\]
and, {using \eqref{eq:CovFourierModesWSpaceTime} and the hermitian symmetry}, the second-order moments
\[
\E\left[\left|\int_{s=t_{m}}^{t_{m+1}} e^{-\frac{t_{m+1}-s}{T_\bk}}\widehat{\rd W}_p(s,\bk)\right|^2\right]=L^3_{\text{\tiny{tot}}}\int_{s=t_m}^{t_{m+1}} e^{-\frac{2(t_{m+1}-s)}{T_\bk}} ds=\frac{L^3_{\text{\tiny{tot}}}T_\bk\left(1-e^{-\frac{2\delta t}{T_\bk}}\right)}{2}.
\]
For any $m$, consider a Gaussian white noise {vector field} $\Gamma_{p,m}(\bx)$ and define the associated Fourier modes $\widehat{\Gamma}_{p,m}(\bk)$ {on a periodic domain:}
\[
\widehat{\Gamma}_{p,m}(\bk)=\int_{[-L_{\text{\tiny{tot}}}/2\ ;\ L_{\text{\tiny{tot}}}/2]^3}e^{-2i\pi \bk \cdot \bx}\Gamma_{p,m}(\bx) \rd \bx \ .
\]
As a consequence of \eqref{eq:CovFourierModesWSpaceTime}, one has the equality in distribution
\begin{equation}
\int_{s=t_{m}}^{t_{m+1}} e^{-\frac{t_{m+1}-s}{T_\bk}}\widehat{\rd W}_p(s,\bk)=\sqrt{\frac{T_\bk\left(1-e^{-\frac{2\delta t}{T_\bk}}\right)}{2}}\widehat{\Gamma}_{p,m+1}(\bk),
\end{equation}
meaning that the random variables on the left and the on the right-hand sides of the identity above are Gaussian, centered, and have the same covariance structure.

This suggests to introduce the numerical scheme defined by
\begin{equation}
\widehat{u}^{(1),\nu}_{i,m+1}(\bk) = e^{-\frac{\delta t}{T_\bk}}\widehat{u}^{(1),\nu}_{i,m}(\bk) + \sqrt{\frac{E_\nu^{\text{\tiny{E}}}(\bk)\left(1-e^{-\frac{2\delta t}{T_\bk}}\right)}{2}}\widehat{P}_{ip}(\bk)\widehat{\Gamma}_{p,m+1}(\bk),
\end{equation}
which can be simulated in practice by sampling a new Gaussian white noise {vector field} and its Fourier modes at each iteration. This means that the numerical approximation defines a discrete time Markov process.

If the initial value $\widehat{u}^{(1),\nu}_{i,0}(\bk)$ is equal in distribution to the initial value $\widehat{u}^{(1),\nu}_{i}(0,\bk)$ of \eqref{eq:SolutionGeneralOUTkFourierModesGeneral}, then for all $m\ge 0$ one has $\widehat{u}^{(1),\nu}_{i,m}(\bk)=\widehat{u}^{(1),\nu}_{i}(t_m,\bk)$ in distribution. As a result, the proposed numerical scheme solves exactly in distribution the Ornstein--Uhlenbeck dynamics \eqref{eq:GeneralOUTkFourierModes} (at grid times $t_m=m\delta t$).

A stationary version of the numerical approximation is obtained by choosing the initial value $\widehat{u}^{(1),\nu}_{i,0}(\bk)$ distributed as the fractional Gaussian vector field studied in subsection \ref{Sec:DefFGF}, which can be sampled following \eqref{eq:DefFMfGv} as
\begin{equation}
\widehat{u}^{(1),\nu}_{i,0}(\bk)=\sqrt{\frac{E_\nu^{\text{\tiny{E}}}(\bk)}{2}}\widehat{P}_{ip}(\bk)\widehat{\Gamma}_{p,0}(\bk).
\end{equation}
It is then straightforward to check that by construction the distribution of $\widehat{u}^{(1),\nu}_{i,m}(\bk)$ is independent of $m$.

Applying a numerical scheme which is exact in distribution for the Ornstein--Uhlenbeck dynamics naturally preserves its stationary Gaussian distribution, this is fundamental to obtain numerical simulations which exhibit the correct spatial regularity and statistical properties.

\subsubsection{Numerical scheme for the differentiable-in-time process ($N\ge 2$)}

The objective of this section is to propose a scheme for the process studied in \ref{Sec:GeneNLayers} for arbitrary $N\ge 2$. Like in the case $N=1$ the aim is to define a sequence of random variables $\left(\widehat{u}_{i,m}^{(N),\nu}(\bk)\right)_{m\ge 0}$ such that one has the equality in distribution $\widehat{u}_{i,m}^{(N),\nu}(\bk)=\widehat{u}_{i}^{(N),\nu}(t_m,\bk)$ for any nonnegative integer $m\ge 0$. {When $N\ge 2$, the construction is more involved since the dynamics is  Markovian only when considering the evolution of the additional underlying layers.}

A Markovian formulation of the dynamics is obtained by considering the unknown
\begin{equation}\label{eq:DefFTotalMarkov}
\widehat{\mathbf{F}}_i^{(N),\nu}(t,\bk)=
\begin{pmatrix}
\widehat{u}_i^{(N),\nu}(t,\bk)\\
\widehat{f}_i^{(N-1),\nu}(t,\bk)\\
\vdots\\
\widehat{f}_i^{(2),\nu}(t,\bk)\\
\widehat{f}_i^{(1),\nu}(t,\bk)
\end{pmatrix} \ .
\end{equation}
Introducing the square matrix $\mathbf{A}^{(N)}(\bk)$ defined by
\begin{equation}\label{eq:ANofk}
\mathbf{A}^{(N)}(k)=
-\frac{\sqrt{4N}}{T_k}
\begin{pmatrix}
1 & 0 & \ldots & 0 & 0 \\
0 & 1 & \ddots & \ddots & 0 \\
\vdots & \ddots & \ddots & \ddots & \vdots \\
0 & \ddots & \ddots & 1 & 0 \\
0 & 0 & \ldots & 0 & 1
\end{pmatrix}
+
\begin{pmatrix}
0 & 1 & \ldots & 0 & 0 \\
0 & 0 & \ddots & \ddots & 0 \\
\vdots & \ddots & \ddots & \ddots & \vdots \\
0 & \ddots & \ddots & 0 & 1 \\
0 & 0 & \ldots & 0 & 0
\end{pmatrix}
\end{equation}
and the {column} vector
\begin{equation}
\mathbf{e}^{(N)}=
\begin{pmatrix}
0\\
\vdots\\
0\\
1
\end{pmatrix},
\end{equation}
the vector-valued unknown $\widehat{\mathbf{F}}_i^{(N),\nu}(t,\bk)$ can be interpreted as the solution to the multi-dimensional  Ornstein--Uhlenbeck dynamics
\begin{equation}\label{eq:MarkEvolFullF}
\rd\widehat{\mathbf{F}}_i^{(N),\nu}(t,\bk)=\mathbf{A}^{(N)}(\bk)\widehat{\mathbf{F}}_i^{(N),\nu}(t,\bk)\rd t+\sqrt{q^{(N)}E_\nu(k)}\widehat{P}_{ip}(\bk)\mathbf{e}^{(N)}\rd\widehat{W}_{p}(t,\bk).
\end{equation}

Considering the matrix exponential $e^{t\mathbf{A}^{(N)}(k)}$ for all $t\ge 0$, the variation of constants formula provides an expression for the solution at any time $t\ge 0$
\begin{equation}
\widehat{\mathbf{F}}_i^{(N),\nu}(t,\bk)=e^{t\mathbf{A}^{(N)}(k)}\widehat{\mathbf{F}}_i^{(N),\nu}(0,\bk)+\sqrt{q^{(N)}E_\nu(k)}\widehat{P}_{ip}(\bk)\!\int\limits_{s=0}^{t}e^{(t-s)\mathbf{A}^{(N)}(k)}\mathbf{e}^{(N)}\rd\widehat{W}_{p}(s,\bk).
\end{equation}

Using \eqref{eq:ANofk}, it is straightforward to check that for all $t\ge 0$ one has
\begin{equation}
e^{t\mathbf{A}^{(N)}(k)}=e^{-\frac{\sqrt{4N} t}{T_\bk}}
\begin{pmatrix}
1 & t & \ldots & \frac{t^{N-2}}{(N-2)!} & \frac{t^{N-1}}{(N-1)!} \\
0 & 1 & \ddots & \ddots & \frac{t^{N-2}}{(N-2)!} \\
\vdots & \ddots & \ddots & \ddots & \vdots \\
0 & \ddots & \ddots & 1 & t \\
0 & 0 & \ldots & 0 & 1
\end{pmatrix}
\end{equation}
and as a result one has
\begin{equation}
e^{t\mathbf{A}^{(N)}(k)}\mathbf{e}^{(N)}
=
e^{-\frac{\sqrt{4N} t}{T_k}}
\begin{pmatrix}
\frac{t^{N-1}}{(N-1)!} \\
\frac{t^{N-2}}{(N-2)!} \\
\vdots \\
t \\
1
\end{pmatrix}
.
\end{equation}
Like in the derivation of the numerical scheme when $N=1$, it is convenient to express the solution at time $t_{m+1}=t_m+\delta t$ depending on the solution at time $t_m$, as
\begin{equation}
\begin{aligned}
\widehat{\mathbf{F}}_i^{(N),\nu}(t_{m+1},\bk)&=e^{\delta t\mathbf{A}^{(N)}(k)}\widehat{\mathbf{F}}_i^{(N),\nu}(t_m,\bk)\\
&+\sqrt{q^{(N)}E_\nu(|\bk|)}\widehat{P}_{ip}(\bk)\int_{s=t_{m}}^{t_{m+1}}e^{(t_{m+1}-s)\mathbf{A}^{(N)}(k)}\mathbf{e}^{(N)}\rd\widehat{W}_{p}(s,\bk).
\end{aligned}
\end{equation}
It is required to sample at each iteration the Gaussian random vector 
\begin{equation}
\widehat{\mathbf{Z}}_{p,m}^{(N)}(\bk)
=
\begin{pmatrix}
\widehat{Z}_{p,m}^{(N)}(\bk)\\
\widehat{Z}_{p,m}^{(N-1)}(\bk)\\
\vdots\\
\widehat{Z}_{p,m}^{(2)}(\bk)\\
\widehat{Z}_{p,m}^{(1)}(\bk)
\end{pmatrix}
=
\int_{s=t_{m}}^{t_{m+1}}e^{(t_{m+1}-s)\mathbf{A}^{(N)}(k)}\mathbf{e}^{(N)}\rd\widehat{W}_{p}(s,\bk)
\end{equation}
which has components given by
\begin{equation}
\widehat{Z}_{p,m}^{(n)}(\bk)=\int_{s=t_{m}}^{t_{m+1}}e^{-\frac{\sqrt{4N}(t_{m+1}-s)}{T_\bk}}\frac{(t_{m+1}-s)^{n-1}}{(n-1)!}\rd\widehat{W}_{p}(s,\bk).
\end{equation}
The components of $\widehat{Z}_{p,m}^{(n)}(\bk)$ are not independent since they depend on the same white noise. In order to sample the centered Gaussian random vector $\widehat{\mathbf{Z}}_{p,m}^{(N)}(\bk)$, one needs to compute its covariance matrix $\mathbf{Q}^{(N)}(\bk)$ with entries given by
\begin{align*}
\left(\mathbf{Q}^{(N)}(\bk)\right)_{n_1,n_2}&=\E[\widehat{Z}_{p,m}^{(n_1)}(\bk)\overline{\widehat{Z}_{p,m}^{(n_2)}(\bk)}]\\
&=L^3_{\text{\tiny{tot}}}\int_{s=t_{m}}^{t_{m+1}}e^{-2\frac{\sqrt{4N}(t_{m+1}-s)}{T_\bk}}\frac{(t_{m+1}-s)^{n_1-1}}{(n_1-1)!} \frac{(t_{m+1}-s)^{n_2-1}}{(n_2-1)!}ds\\
&=L^3_{\text{\tiny{tot}}}\int_{s=0}^{\delta t}e^{-2\frac{\sqrt{4N}s}{T_\bk}}\frac{s^{n_1-1}}{(n_1-1)!} \frac{s^{n_2-1}}{(n_2-1)!}ds.
\end{align*}
In order to sample a centered $N$-dimensional Gaussian random vector with a given covariance matrix $\mathbf{Q}$, it is sufficient to identify any matrix $\mathbf{R}$ such that $\mathbf{Q}=\mathbf{R}\mathbf{R}^T$. Indeed, if $\Gamma$ is a Gaussian random vector with independent components distributed as standard centered normal random variables, then $\mathbf{R}\Gamma$ is a centered Gaussian random vector with covariance matrix $\mathbf{R}\mathbf{R}^T=\mathbf{Q}$. The identification of such matrices $\mathbf{R}$ can be performed in several ways. Recalling that the covariance matrix $\mathbf{Q}$ is nonnegative symmetric, one may compute its square root (which requires to compute first its eigenvalues and eigenvectors). In practice, when $\mathbf{Q}$ is positive definite, a more convenient approach is to compute the Cholesky decomposition of $\mathbf{Q}$, which amounts to finding a matrix $\mathbf{R}$ that is lower triangular. The matrix $\mathbf{R}$ is unique if the diagonal entries of $\mathbf{Q}$ are positive. Moreover, Cholesky decomposition can be computed using a straightforward algorithm.

Let $\mathbf{R}^{(N)}(\bk)$ denote the matrix obtained by the Cholesky decomposition of the covariance matrix $\mathbf{Q}^{(N)}(\bk)$ whose entries are given above. The Gaussian random vector $\widehat{\mathbf{Z}}_{p,m}^{(N)}(\bk)$ can be sampled as $\mathbf{R}^{(N)}(\bk)\widehat{\mathbf{\Gamma}}_{p,m}^{(N)}$,
where $\widehat{\mathbf{\Gamma}}_{p,m}^{(N)}$ is a Gaussian random vector given by
\[
\widehat{\mathbf{\Gamma}}_{p,m}^{(N)}=\begin{pmatrix}
\widehat{\Gamma}_{p,m}^{(N)}(\bk)\\
\widehat{\Gamma}_{p,m}^{(N-1)}(\bk)\\
\vdots\\
\widehat{\Gamma}_{p,m}^{(2)}(\bk)\\
\widehat{\Gamma}_{p,m}^{(1)}(\bk)
\end{pmatrix}
\]
with entries $\widehat{\Gamma}_{p,m}^{(N)}(\bk),\ldots,\widehat{\Gamma}_{p,m}^{(1)}(\bk)$ given as Fourier modes of independent Gaussian white noise {vector fields} $\Gamma_{p,m}^{(N)}(\bx),\ldots,\Gamma_{p,m}^{(1)}(\bx)$.

Finally, one obtains the numerical scheme defined by
\begin{equation}
\widehat{\mathbf{F}}_{i,m+1}^{(N),\nu}(\bk)=e^{\delta t\mathbf{A}^{(N)}(k)}\widehat{\mathbf{F}}_{i,m}^{(N),\nu}(\bk)+\sqrt{q^{(N)}E_\nu(|\bk|)}\widehat{P}_{ip}(\bk)\mathbf{R}^{(N)}(\bk)\widehat{\mathbf{\Gamma}}_{p,m}^{(N)}.
\end{equation}
The numerical approximation defines a discrete time Markov process, where at each iteration $N$ new Gaussian white noise {vector fields} and their Fourier modes need to be computed.

If the initial value $\widehat{\mathbf{F}}_{i,0}^{(N),\nu}(\bk)$ is equal in distribution to the initial value $\widehat{\mathbf{F}}_{i}^{(N),\nu}(0,\bk)$ of the continuous time dynamics, then for all $m\ge 0$ one has $\widehat{\mathbf{F}}_{i,m}^{(N),\nu}(\bk)=\widehat{\mathbf{F}}_{i}^{(N),\nu}(t_m,\bk)$ in distribution. As a result, the proposed numerical scheme solves exactly in distribution the Ornstein--Uhlenbeck dynamics (at grid times $t_m=m\delta t$).

A stationary version of the numerical approximation is obtained by choosing the initial value $\widehat{\mathbf{F}}_{i,0}^{(N),\nu}(\bk)$ distributed according to the Gaussian stationary distribution described in subsection~\ref{Sec:GeneNLayers}. If this holds, then the distribution of $\widehat{\mathbf{F}}_{i,m}^{(N),\nu}(\bk)$ is independent of $m$.

\section{Conclusions and perspectives}\label{Sec:Conclusion}

We have proposed a time evolving stochastic model of a Gaussian velocity field that reproduces several important aspects of   fluid turbulence, up to second-order statistical quantities. The present version of the model is formulated in a Gaussian framework and is fully characterized by its spectral properties, or equivalently by its second-order structure function, in both space and time. To the best of our knowledge, this is the unique stochastic representation of fluid turbulence with a realistic statistical temporal structure, associated to a {Markovian} formulation, which encompasses the dimensional predictions of \cite{Ten75}. The statistical signatures of this random field compare in a satisfactory manner with the observations made on DNS velocity fields,   provided by the Johns Hopkins database \citep{LiPer08}. We provide efficient numerical schemes, exact in probability distributions, that are straightforward to implement.

As mentioned throughout the article, the present random model is a Gaussian field and therefore misses the possibility of reproducing several other fluid turbulence properties of crucial importance. Among them, the model misses the asymmetrical nature of the probability density functions of the velocity increments, known in turbulence as the skewness phenomenon \citep{Fri95}. The skewness phenomenon is clearly understood in the context of energy transfers. Its statistical formulation is fully taken into account by the Kármán and Howarth equation \citep{Fri95} and even locally in the context of the budget of weak solutions of the Navier-Stokes equations \citep{DucRob00,Eyi24}. It is usually associated with the existence of a cascading phenomenon of energy towards small scales. Thus, in this sense, the present Gaussian model is not able to reproduce some crucial ingredients of the nonlinear dynamics, that are at the heart of the equations of fluid motion. In the same spirit, other non-Gaussian behaviors, as can be observed in higher-order statistical quantities, such as the flatness of velocity increments \citep{CheCas12}, are not reproduced either. These anomalous behaviors are usually associated to the intermittency phenomenon, also known as multifractality \citep{Fri95}. In a future communication (part II), we aim at generalizing the present Gaussian approach to a multifractal framework, while including some ingredients of the vorticity stretching phenomenon, in the spirit of previous works \citep{CheRob10,PerGar16}.

Also, as we mentioned in Section \ref{Sec:CompTempStatDNSModel} when we compared the temporal structure of the present modeled velocity field with DNS, the present approach misses an important ingredient of the sweeping effect, but presents excellent agreement from a statistical point of view. A recent proposition by \cite{ArmVic25} seems to allow this property by introducing in their iterative construction of a velocity field the inverse of the Lagrangian flow map. It is thus tempting to apply such a procedure in the context of fractional Gaussian fields and explore whether it gives a realistic picture in time and if this remains affordable from a numerical point of view. {We believe that this realistic picture of the temporal structure would have tremendous importance when such a velocity field advects a given scalar in a passive \citep{FalGaw01} or active \citep{JosAwa25} way,  with direct implications on the regularity of velocity along Lagrangian trajectories \citep{RenChe20}.} Once again, we keep these aspects in mind for future investigations.

\smallskip 
\noindent \backsection[Acknowledgements]{L.C. and I. S. are partially supported and J. D. L. is funded by the Simons Foundation Award ID: 651475. L.C. is also funded by
Agence National de la Recherche ANR-20-CE30-0035. We thank Mokhtar Adda Bedia for his help regarding the expansion of Bessel functions in the vicinity of the origin. This work is also partially supported by Institut des Mathématiques pour la Planète Terre (Project: The role of turbulent fluctuations in the natural dispersion of aerosols, particles and pollutants).}

\noindent \backsection[Declaration of Interests]{The authors report no conflict of interest.}

\noindent \backsection[Note added in proof]{We have become aware of the existence of a recent preprint by \cite{AwaJos26} where is considered the one-layer velocity field $\bu^{(1),\nu}(t,\bx)$ which Fourier mode evolution is provided in \eqref{eq:GeneralOUTkFourierModes}}

\appendix

\section{Analysis of the temporal covariance structure of the embedded model}\label{App:CovEmbeddedModel}

\subsection{Computation and limit of the covariance structure of $\widehat{u}_i^{(N),\nu}(t,\bk_\bn)$ as $N\to\infty$. }

Following Proposition A.1 of Ref. \cite{VigFri20}, we can compute the covariance function of $\widehat{u}^{(N),\nu}_i(t,\bk_\bn)$. Let us reproduce here the arguments developed in \cite{VigFri20}. Let us first assume that the system of $N$ equations \eqref{eq:OU_embedded_N_1} to \eqref{eq:OU_embedded_N_N} possesses a statistically stationary solution, with the consequence that correlation functions only depend on the time difference, such that 
\begin{align}
\mathcal C_{\widehat{u}_i^{(N),\nu},\widehat{u}_j^{(N),\nu}}(\tau,\bk,\bk')=\E\left[\widehat{u}^{(N),\nu}_i(t,\bk)\widehat{u}^{(N),\nu}_j(t+\tau,\bk')\right],
\end{align}
and for any $1\le n\le N-1$,
\begin{align}
\mathcal C_{\widehat{f}_i^{(n),\nu},\widehat{f}_j^{(n),\nu}}(\tau,\bk,\bk')=\E\left[\widehat{f}^{(n),\nu}_i(t,\bk)\widehat{f}^{(n),\nu}_j(t+\tau,\bk')\right].
\end{align}

In the statistically stationary regime, we can write, similarly to \eqref{eq:SolutionStatStatOUTkFourierModes},
\begin{align}
\widehat{u}^{(N),\nu}_i(t,\bk_\bn)=\int_{-\infty}^{t} \; e^{-\sqrt{4N}(t-s)/T_{\bk_\bn}}\widehat{f}^{(N-1),\nu}_i(s,\bk_\bn)\rd s,
\end{align}
and as a result, one obtains
\begin{align}
\mathcal C_{\widehat{u}_i^{(N),\nu},\widehat{u}_j^{(N),\nu}}&(\tau,\bk_\bn,\bk_\bm)\notag\\
&=\int_{-\infty}^{t}\int_{-\infty}^{t+\tau} \; e^{-\sqrt{4N}(2t+\tau-s_1-s_2)/T_{\bk_\bn}}\mathcal C_{\widehat{f}_i^{(N-1),\nu},\widehat{f}_j^{(N-1),\nu}}(s_1-s_2,\bk_\bn,\bk_\bm)\rd s_1\rd s_2\notag\\
&=\int_{-\infty}^{0}\int_{-\infty}^{\tau} \; e^{-\sqrt{4N}(\tau-s_1-s_2)/T_{\bk_\bn}}\mathcal C_{\widehat{f}_i^{(N-1),\nu},\widehat{f}_j^{(N-1),\nu}}(s_1-s_2,\bk_\bn,\bk_\bm)\rd s_1\rd s_2\notag\\
&=\int_{0}^{\infty}\int_{-\tau}^{\infty} \; e^{-\sqrt{4N}(\tau+s_1+s_2)/T_{\bk_\bn}}\mathcal C_{\widehat{f}_i^{(N-1),\nu},\widehat{f}_j^{(N-1),\nu}}(s_1-s_2,\bk_\bn,\bk_\bm)\rd s_1\rd s_2,
\end{align}
where the last equality follows by the parity in time of any correlation function. The above can be formally rewritten as 
 \begin{align*}%\label{eq:corrvndef}
 \mathcal C_{\widehat{u}_i^{(N),\nu},\widehat{u}_j^{(N),\nu}}(\tau,\bk_\bn,\bk_\bm)&= \int_{\mathbb R^2} g_{T_{\bk_\bn}}(\tau+s_2)g_{T_{\bk_\bn}}(s_1)\mathcal C_{\widehat{f}_i^{(N-1),\nu},\widehat{f}_j^{(N-1),\nu}}(s_1-s_2,\bk_\bn,\bk_\bm)\rd s_1\rd s_2\\
 &=\int_{\mathbb R^2} g_{T_{\bk_\bn}}(\tau+t_1+t_2)g_{T_{\bk_\bn}}(t_1)\mathcal C_{\widehat{f}_i^{(N-1),\nu},\widehat{f}_j^{(N-1),\nu}}(t_2,\bk_\bn,\bk_\bm)\rd t_1\rd t_2\\
 &=\int_{\mathbb R} \left(g_{T_{\bk_\bn}}\star g_{T_{\bk_\bn}}\right)(\tau+t_2)\mathcal C_{\widehat{f}_i^{(N-1),\nu},\widehat{f}_j^{(N-1),\nu}}(t_2,\bk_\bn,\bk_\bm))\rd t_2\\
 &=\left(g_{T_{\bk_\bn}}\star g_{T_{\bk_\bn}}\star \mathcal C_{\widehat{f}_i^{(N-1),\nu},\widehat{f}_j^{(N-1),\nu}}(\cdot ,\bk_\bn,\bk_\bm)\right)(\tau),
\end{align*}
where we have introduced the function $g_{T_{\bk}}(t)=e^{-\sqrt{4N}t/T_{\bk}}\mathbbm{1}_{t\ge 0}$ and the correlation product $\star$, which is defined, for any two functions $g_1$ and $g_2$, as
$$ \left(g_1\star g_2\right)(\tau) = \int_{\mathbb R}g_1(t)g_2(t+\tau)\rd t.$$
Note now that the correlation product of $g_{T_\bk}$ by itself is given by
\begin{equation}\label{eq:TransFoncOU}
G_{T_\bk}(t) \equiv \left(g_{T_\bk}\star g_{T_\bk}\right)(t) = \frac{T_\bk}{2\sqrt{4N}}e^{-\sqrt{4N}|t|/T_\bk}.
\end{equation} 
Therefore, by induction, we obtain
 \begin{align*}%\label{eq:corrvndef}
 \mathcal C_{\widehat{u}_i^{(N),\nu},\widehat{u}_j^{(N),\nu}}(\tau,\bk_\bn,\bk_\bm)&=\left(G_{T_{\bk_\bn}}^{\star (N-1)}\star \mathcal C_{\widehat{f}_i^{(1),\nu},\widehat{f}_j^{(1),\nu}}(\cdot ,\bk_\bn,\bk_\bm)\right)(\tau),
\end{align*}
using the  notation
$$g^{\star n}=\underbrace{g \star g \star \cdot \cdot \cdot \star g}_{n}.$$
Finally, noting that 
$$ \mathcal C_{\widehat{f}_i^{(1),\nu},\widehat{f}_j^{(1),\nu}}(\cdot ,\bk_\bn,\bk_\bm) = q^{(N)}L^3_{\text{\tiny{tot}}} G_{T_{\bk_\bn}}(\tau)E_\nu^{\text{\tiny{E}}}(k_\bn)\widehat{P}_{ij}(\bk_\bn)\delta_{\bn,-\bm}^{(3)},$$
we obtain the expression of the temporal correlation function 
 \begin{align}\label{eq:corrhatUN}
 \mathcal C_{\widehat{u}_i^{(N),\nu},\widehat{u}_j^{(N),\nu}}(\tau,\bk_\bn,\bk_\bm)= q^{(N)}L^3_{\text{\tiny{tot}}} G_{T_{\bk_\bn}}^{\star N}(\tau)E_\nu^{\text{\tiny{E}}}(k_\bn)\widehat{P}_{ij}(\bk_\bn)\delta_{\bn,-\bm}^{(3)}.
\end{align}

In order to study the asymptotic behavior when $N\to \infty$, the function $G_{T_\bk}(t)$ \eqref{eq:TransFoncOU} can be written in the frequency domain as
\begin{equation}\label{eq:GTkFreqDom}
G_{T_\bk}(\tau) = \int_{\mathbb R}e^{2i\pi \omega\tau}\frac{\frac{T_\bk^{2}}{4N}}{1+\pi^2\frac{T_\bk^{2}}{N}\omega^2}d\omega,
\end{equation} 
and one obtains
\begin{equation}\label{GTkFreqDomStarN}
G_{T_\bk}^{\star N}(\tau) = \int_{\mathbb R}e^{2i\pi \omega\tau}\left[\frac{\frac{T_\bk^{2}}{4N}}{1+\pi^2\frac{T_\bk^{2}}{N}\omega^2} \right]^N d\omega.
\end{equation}
Taking into account the value of $q^{(N)}$ \eqref{eq:ValueqN}, we obtain
 \begin{align}\label{eq:corrhatUNSpect}
 \mathcal C_{\widehat{u}_i^{(N),\nu},\widehat{u}_j^{(N),\nu}}(\tau,\bk_\bn,\bk_\bm)&= \frac{L^3_{\text{\tiny{tot}}}}{2}F^{(N)}(\tau/T_{\bk_\bn})E_\nu^{\text{\tiny{E}}}(k_\bn)\widehat{P}_{ij}(\bk_\bn)\delta_{\bn,-\bm}^{(3)},
 \end{align}
 with  the kernel $F^{(N)}$ given by
 \begin{align}\label{eq:ExprFNIntegralst}
F^{(N)}(\tau)=\frac{\int_{\mathbb R}e^{2i\pi \omega\tau}\frac{1}{\left[1+\frac{\pi^2\omega^2}{N}\right]^N} d\omega}{ \int_{\mathbb R}\frac{1}{\left[1+\frac{\pi^2\omega^2}{N}\right]^N} d\omega}.
 \end{align}
 Note that the expression provided in \eqref{eq:CovHatUFiniteN} with the temporal kernel $F^{(N)}(\tau/T_\bk)$ given in \eqref{eq:FNTAU} follows from \eqref{eq:corrhatUNSpect} once the remaining integrals entering in \eqref{eq:ExprFNIntegralst} are computed with the help of a symbolic calculation software that also allows us to give the explicit expression in \eqref{eq:ValueqN}. 
 
 Unlike the Fourier mode $\widehat{u}_i^{(1),\nu}(t,\cdot )$ which is not differentiable, the Fourier mode $\widehat{u}_i^{(N),\nu}(t,\cdot )$ for $N\ge 2$ is now differentiable (N-1) times. In particular, the variance of its derivatives $\rd^p\widehat{u}_i^{(N),\nu}(t,\bk_\bn)/\rd t^p$ becomes finite as long as $p\le N-1$, and we can get the expression
 \begin{align}\label{eq:ExprePDerivativesUN}
 \E\left[\left| \frac{\rd ^p\widehat{u}^{(N),\nu}(t,\bk_\bn)}{\rd t^p}\right|^2\right]&=L^3_{\text{\tiny{tot}}}E_\nu^{\text{\tiny{E}}}(k_\bn)(-1)^p\left.\frac{\rd ^{2p}F^{(N)}(\tau/T_{\bk_\bn})}{\rd t^{2p}}\right|_{\tau=0}\notag\\
 &=L^3_{\text{\tiny{tot}}}E_\nu^{\text{\tiny{E}}}(k_\bn)\frac{(2\pi)^{2p}}{T_{\bk_\bn}^{2p}}\frac{\int_{\mathbb R}\frac{\omega^{2p}}{\left[1+\frac{\pi^2\omega^2}{N}\right]^N} d\omega}{ \int_{\mathbb R}\frac{1}{\left[1+\frac{\pi^2\omega^2}{N}\right]^N} d\omega}
 \end{align}
which can be shown to be finite as long as $p\le N-1$, and can eventually be exactly expressed.

 More interesting is the limit of \eqref{eq:ExprFNIntegralst} when $N\to \infty$. Using the expression of the exponential as the limit
 \begin{align}
 \lim_{N\to\infty} \frac{1}{\left( 1+\frac{x}{N}\right)^N} = e^{-x},
 \end{align} 
 permutation of the integrals and the limits ensured by a dominated convergence argument \citep{VigFri20},  one then obtains   \eqref{eq:ConvTowardGaussianKernel} following from \eqref{eq:ExprFNIntegralst} in the limit $N\to\infty$.

\subsection{Computation of the equivalent of the kernel $F^{(N)}$ at small argument }

We will eventually need the behavior of the function $F^{(N)}(\tau)$ at small argument $\tau\to 0$. An analytical expression of \eqref{eq:ExprFNIntegralst} can be computed using Mathematica which reads:
 \begin{align}\label{eq:FNTAUAnnex}
 F^{(N)}(\tau) = \frac{2 |\sqrt{N}\tau|^{N-1/2}
 K_{N-1/2}\left( 2|\sqrt{N}\tau|\right)}{\Gamma(N-1/2)}.
 \end{align}
To do so, we have to perform an {expansion} of the modified Bessel function of the second kind. We found the approach developed by \cite{SteKat24} very clear and convenient. Based on their equations (2) and (3), we write
$$K_{N-1/2}\left( 2|\sqrt{N}\tau|\right)=(-1)^N\frac{\pi}{2}\left[I_{N-1/2}\left( 2|\sqrt{N}\tau|\right)-I_{-N+1/2}\left( 2|\sqrt{N}\tau|\right) \right],$$
which depends on the modified Bessel function of the first kind $I_\alpha(x)$ with the following {expansion} 
$$ I_\alpha(x) =\left( \frac{x}{2}\right)^\alpha\sum_{k=0}^\infty \frac{1}{4^k k! \Gamma(\alpha+k+1)}x^{2k}=\left( \frac{x}{2}\right)^\alpha\left( \frac{1}{\Gamma(\alpha+1)}+\frac{1}{4 \Gamma(\alpha+2)}x^{2}+o(x^2)\right).$$
At small argument, we get the following
\begin{align}\label{eq:FNTAUAnnexDevel}
 F^{(N)}(\tau) = (-1)^{N+1}\pi \frac{1}{\Gamma(N-1/2)}\left( \frac{1}{\Gamma(-N+3/2)}+\frac{N\tau^{2}}{ \Gamma(-N+5/2)}+o(\tau^2)\right),
 \end{align}
which, using
$$ \Gamma(-N+3/2)\Gamma(N-1/2) = \frac{\pi}{\sin(\pi (-N+3/2))}=\pi(-1)^{N+1}$$
and
$$ \Gamma(-N+5/2)\Gamma(N-1/2) =(-N+3/2)\pi(-1)^{N+1},$$
simplifies to,
\begin{align}\label{eq:FNTAUAnnexDevelSimp}
 F^{(N)}(\tau) = 1-\frac{N}{ N-3/2}\tau^{2}+o(\tau^2).
 \end{align}
 
{ \section{Derivation of the temporal structure function and frequency spectrum in the multi-layered framework}}\label{App:S2andEMultiLayered}
 
 {Let us now explore the consequences of the particular temporal kernel $F^{(N)}$ given in \eqref{eq:FNTAU} (resp. $F^{(\infty)}$ given in \eqref{eq:FInfiniteNTAU}) on the regularity in time of the corresponding velocity field $u^{(N),\nu}_i(t,\bx)$ (resp. $\bu^{(\infty),\nu}_i(t,\bx)$).
This will be illustrated with the scaling behavior of the temporal structure function, i.e. the variance of the velocity time increment \eqref{eq:TimeIncrement}. Using the generalized forms of $ E^{\text{\tiny{E}}}(k)$ \eqref{eq:ModelPSDAnyHNuTo0} and $T_k$, \eqref{eq:DefTkBeta}, the expression \eqref{eq:TemporalS2} reads 
\begin{align}\label{eq:TemporalS2AnyHFN}
S_2^{(N),\text{\tiny{temp}}}(\tau)&=\lim_{L_{\text{\tiny{tot}}}\to\infty}\lim_{\nu\to 0}\E\left[ \left|\delta_\tau \bu^{(N),\nu}\right|^2\right]\notag\\
&= \frac{1}{\pi}(1+2H)(3+2H) D_2 \int_{\bk \in \R^3} \left[1-F^{(N)}\left(D_3|\bk|_{1/L}^{2\beta}|\tau|\right) \right]|\bk|^2 |\bk|_{1/L}^{-(2H+5)} \ \rd \bk ,
\end{align}
and when $N\to\infty$
\begin{align}\label{eq:TemporalS2AnyHGauss}
S_2^{(\infty),\text{\tiny{temp}}}(\tau)&=\lim_{L_{\text{\tiny{tot}}}\to\infty}\lim_{\nu\to 0}\E\left[ \left|\delta_\tau \bu^{(\infty),\nu}\right|^2\right]\notag\\
&= \frac{1}{\pi}(1+2H)(3+2H) D_2 \int_{\bk \in \R^3} \left[1-F^{(\infty)}\left(D_3|\bk|_{1/L}^{2\beta}|\tau|\right) \right]|\bk|^2 |\bk|_{1/L}^{-(2H+5)} \ \rd \bk .
\end{align}}

{Similarly to the temporal structure function \eqref{eq:TemporalS2AnyH} associated with the exponential kernel \eqref{eq:FTempCorrOU}, $S_2^{(N),\text{\tiny{temp}}}(\tau)$ \eqref{eq:TemporalS2AnyHFN} and $S_2^{(\infty),\text{\tiny{temp}}}(\tau)$ \eqref{eq:TemporalS2AnyHGauss} eventually undergo a transition at small scales, but for a different range of values of $H$ and $\beta$. We indeed obtain, 
\begin{align}\label{eq:TemporalS2AnyHEquiv1FN}
S_2^{(N),\text{\tiny{temp}}}&(\tau)\build{\sim}_{|\tau|\to 0}^{}\notag\\
&4(1+2H)(3+2H) D_2 |D_3\tau|^{H/\beta}\int_{\rho=0}^{\infty} \left[1-F^{(N)}\left(\rho^{2\beta}\right) \right]\rho^{-(2H+1)} \rd \rho \mbox{ for }H<2\beta.
\end{align}
Similarly for $S_2^{(\infty),\text{\tiny{temp}}}(\tau)$, replacing $F^{(N)}$ by $F^{(\infty)}$ in \eqref{eq:TemporalS2AnyHEquiv1FN} and computing the remaining integral using the Gaussian correlation function \eqref{eq:FInfiniteNTAU}, one obtains
\begin{align} \label{eq:TemporalS2AnyHEquiv1FNInfinite}
S_2^{(\infty),\text{\tiny{temp}}}(\tau)&\build{\sim}_{|\tau|\to 0}^{}4(1+2H)(3+2H) D_2 |D_3\tau|^{H/\beta}\frac{\Gamma\left(1-\frac{H}{2\beta}\right)}{2H}\quad \mbox{ for }H<2\beta.
\end{align}
Note that, for a given $\beta$, the range of accessible spatial regularity $H$ is wider (i.e. $H<2\beta$) in \eqref{eq:TemporalS2AnyHEquiv1FN} and \eqref{eq:TemporalS2AnyHEquiv1FNInfinite} than what was found (i.e. $H<\beta$) in \eqref{eq:TemporalS2AnyHEquiv1}. This is a consequence of the behavior of $1-F^{(N)}(x)$ and $1-F^{(\infty)}(x)$ near the origin which go to zero as $x^2$, while the exponential kernel \eqref{eq:FTempCorrOU} predicts a much slower decay (i.e. as $x$). This is shown in \eqref{eq:FNTAUAnnexDevelSimp}.}

{For a given $\beta$, when $H$ gets large, one obtains the trivial scaling associated to a {differentiable} field, using the expansion of the function $F^{(N)}$ provided in \eqref{eq:FNTAUAnnexDevelSimp}:
\begin{align}\label{eq:TemporalS2AnyHFNTrivial}
S_2^{(N),\text{\tiny{temp}}}&(\tau)\build{\sim}_{|\tau|\to 0}^{} \notag\\
&\frac{1}{\pi}(1+2H)(3+2H) D_2 \left| D_3\tau\right|^2\frac{N}{N-3/2}\int_{\bk \in \R^3}|\bk|^2 |\bk|_{1/L}^{4\beta-(2H+5)} \rd\bk \mbox{ for }H>2\beta.
\end{align}
A similar result is satisfied  for $S_2^{(\infty),\text{\tiny{temp}}}(\tau)$ as a consequence of \eqref{eq:TemporalS2AnyHFNTrivial} when $N\to\infty$.}

{The scaling behaviors derived in \eqref{eq:TemporalS2AnyHEquiv1FN} and \eqref{eq:TemporalS2AnyHFNTrivial} pertain to the inertial range, for which the viscosity has been set to 0 before considering any limiting behavior. We now explore the behavior of the structure function taking first the limit of vanishing scales $\tau\to 0$, and second the limit of vanishing viscosity $\nu\to 0$. When the viscosity $\nu$ is not taken to zero, instead of \eqref{eq:TemporalS2AnyHEquiv1FN}, we have
\begin{align}\label{eq:TemporalS2AnyHViscFN}
S_{2,\nu}^{(N),\text{\tiny{temp}}}(\tau)=\lim_{L_{\text{\tiny{tot}}}\to\infty}\E\left[ \left|\delta_\tau \bu^{(N),\nu}\right|^2\right]&= 2 \int_{\bk \in \R^3} \left[1-F^{(N)}\left(D_3|\bk|_{1/L}^{2\beta}|\tau|\right) \right] E_\nu^{\text{\tiny{E}}}(\bk)\rd \bk\notag\\
&\build{\sim}_{|\tau|\to 0}^{} 2\left| D_3\tau\right|^2\frac{N}{N-3/2}\int_{\bk \in \R^3}|\bk|_{1/L}^{4\beta}E_\nu^{\text{\tiny{E}}}(\bk) \ \rd \bk ,
\end{align}
which behaves proportionally to $\tau^2$ for any pair of values of $H$ and $\beta$. This is very different from the behavior \eqref{eq:TemporalS2AnyHVisc} obtained for the simple exponential kernel \eqref{eq:FTempCorrOU}: the obtained velocity field $\bu^{(N),\nu}(t,\bx)$ is now smooth in time as long as $\nu>0$ and $N\ge 2$.}

{In the case $H<2\beta$, we also observe a transition in the power-law exponent of the temporal structure function, taking value $S_{2}^{(N),\text{\tiny{temp}}}(\tau)\propto |\tau|^{H/\beta}$ in the inertial range \eqref{eq:TemporalS2AnyHEquiv1FN}, and then $S_{2,\nu}^{(N),\text{\tiny{temp}}}(\tau)\propto |\tau|^{2}$ in the dissipative range \eqref{eq:TemporalS2AnyHViscFN}. This transition takes place at a dissipative time scale $\tau_d$ defined as the very characteristic time scale at which the asymptotic expansions of both $S_{2}^{(N),\text{\tiny{temp}}}$ (obtained from \eqref{eq:TemporalS2AnyHEquiv1FN} ) and $S_{2,\nu}^{(N),\text{\tiny{temp}}}$ (obtained from \eqref{eq:TemporalS2AnyHViscFN}) coincide. Then it is clear that $\tau_d$ is fully determined by viscosity and can be shown to be related to a certain power (depending on $\beta$) of the dissipative length scale $\eta_d$ defined in the longitudinal PSD \eqref{eq:ModelPSD1dLongAnyH}.}

{Let us now study the temporal spectrum $E_\nu^{(N),\text{\tiny{T}}}(\omega)$ (resp. $E_\nu^{(\infty),\text{\tiny{T}}}(\omega)$) of the resulting velocity field $\bu^{(N),\nu}$ (resp. $\bu^{(\infty),\nu}$) in a similar way as we did before for $\bu^{(1),\nu}$. We have
\begin{align}\label{eq:TimeSpectrumFN}
E_\nu^{(N),\text{\tiny{T}}}(\omega) &= \lim_{L_{\text{\tiny{tot}}}\to\infty}\int_{\tau\in\R}e^{-2i\pi \omega \tau} \E \left[ \bu^{(N),\nu}(t,\bx)\cdot \bu^{(N),\nu}(t+\tau,\bx)\right]\rd \tau \\
&= \int_{\bk \in \R^3} T_\bk\widehat{F}^{(N)}(\omega T_\bk) E_\nu^{\text{\tiny{E}}}(\bk) \ \rd \bk ,\notag
\end{align}
where  $\widehat{F}^{(N)}$ is the Fourier transform of the temporal kernel $F^{(N)}$ \eqref{eq:FNTAU}. A simple expression is obtained directly from \eqref{eq:ExprFNIntegralst}:
\begin{align}\label{eqFourierFN}
\widehat{F}^{(N)}(\omega)= \frac{\sqrt{\pi}\Gamma(N)}{\sqrt{N}\Gamma(N-1/2)}\frac{1}{\left( 1+\frac{\pi^2\omega^2}{N}\right)^N}.
\end{align}}

{In the limit $N\to\infty$, a similar expression for $E_\nu^{(\infty),\text{\tiny{T}}}(\omega)$ could be obtained, by replacing $\widehat{F}^{(N)}$ in \eqref{eq:TimeSpectrumFN} by the  Fourier transform $\widehat{F}^{(\infty)}$ of the temporal kernel $F^{(\infty)}$ \eqref{eq:FInfiniteNTAU}, which is given by 
\begin{align}\label{eqFourierFInfinity}
\widehat{F}^{(\infty)}(\omega)= \sqrt{\pi}e^{-\pi^2\omega^2},
\end{align}
eventually leading to \eqref{eq:FInfiniteNTAU} after computing the inverse Fourier transform. }

{\subparagraph{}We show in Fig. \ref{Fig:TempCorr1D}(b) the convergence of the 1D Fourier modes to the Gaussian correlation function. Details are explained in the Appendices \ref{App:CovEmbeddedModel} and \ref{App:1DTempCorr} 
Letting first viscosity vanish ($\nu\to 0$), and considering the large frequencies  $\omega\to\infty$ (secondly), we obtain for $H<(2N-1)\beta$
\begin{align}\label{eq:TimeSpectrumFNOmegaInfty}
E^{(N),\text{\tiny{T}}}(\omega) &\equiv \lim_{\nu\to 0} E_\nu^{(N),\text{\tiny{T}}}(\omega)= \int_{\bk \in \R^3} T_{\omega^{\frac{1}{2\beta}} \bk}\widehat{F}^{(N)}(\omega T_{\omega^{\frac{1}{2\beta}} \bk}) E^{\text{\tiny{E}}}(\omega^{\frac{1}{2\beta}} |\bk|) \omega^{\frac{3}{2\beta}} \ \rd \bk ,\notag\\
&\build{\sim}_{\omega\to \infty}^{}2(1+2H)(3+2H) D_2D_3^{H/\beta}\omega^{-\left(1+H/\beta\right)}\int_{0}^\infty \rho^{-\left(2\beta+2H+1 \right)}\widehat{F}^{(N)}(\rho^{-2\beta})\rd \rho.
\end{align}
with the exact expression as $N\to \infty$, without any constraints on the set of parameters $H$ and $\beta$ as long as they are positive,
\begin{align}\label{eq:TimeSpectrumFNInfinityOmegaInfty}
E^{(\infty),\text{\tiny{T}}}(\omega) &\equiv \lim_{\nu\to 0} E_\nu^{(\infty),\text{\tiny{T}}}(\omega)\build{\sim}_{\omega\to \infty}^{}\frac{\sqrt{\pi}}{2\beta}(1+2H)(3+2H) D_2D_3^{H/\beta}(\pi\omega)^{-\left(1+H/\beta\right)} \Gamma\left(\frac{1}{2} + \frac{H}{2\beta}\right).
\end{align}
It is worth noting that, when $N$ is finite,  for $H>(2N-1)\beta$, one has
\begin{align}\label{eq:TimeSpectrumFNOmegaInftyTrivial}
E^{(N),\text{\tiny{T}}}(\omega) &\build{\sim}_{\omega\to \infty}^{}
\frac{1}{2\pi^{2N+1/2}} (1+2H)(3+2H) D_2D_3^{2N-1}N^{N-1/2}\frac{\Gamma(N)}{\Gamma(N-1/2)}\omega^{-2N}\notag\\
&\times \int_{\bk\in\R^3} |\bk|^2|\bk|_{1/L}^{2(2N-1)\beta-(2H+5)}\ \rd \bk .
\end{align}
}

{Equations \eqref{eq:TimeSpectrumFNOmegaInfty} and \eqref{eq:TimeSpectrumFNInfinityOmegaInfty} show that, when $H<2\beta$, even if we let the number of layers $N$ tends to infinity, the field is not differentiable in time, as it is expected in the limit of vanishing viscosity. This range of possible values of the parameter $H$ setting the regularity in time is only accessible when $N\ge 2$. Depending on the values of $N$ and $\beta$,  the present approach allows to consider Hurst parameters $H$ bigger that unity, that gives rise to fields at least once differentiable in time, although physically not really realistic in a turbulent context. When $H$ gets really large, and larger than $(2N-1)\beta$, the equivalent provided in \eqref{eq:TimeSpectrumFNOmegaInftyTrivial} says that the field is differentiable $(N-1)$-times, the derivatives of order $N-1$ having the regularity of the Brownian motion.}

\section{Relationship between the time-correlation of the three-dimensional Fourier modes and the one-dimension longitudinal Fourier modes}\label{App:Rel3d1dTimeCorrelation}
While most of the experimental and numerical statistics are expressed in terms of the longitudinal spectrum  and longitudinal Fourier modes \cf \ \citep{GorBal21}, the present model is made to first prescribe the 3D PSD $E_\nu^{\text{\tiny{E}}}\lp k \rp$. We therefore compute a relation between the longitudinal Fourier modes \cf \ and the 3D Fourier modes \cf . 
\begin{equation}
\begin{aligned}
 C_{\text{\tiny{long}}}^{(N),\nu} \lp \tau, k_x\rp & := \mathbb{E}\left[ \hat{u}_{x}^{(N),\nu}\lp t+\tau, k_x, y, z\rp \overline{\hat{u}_{x}^{(N),\nu}\lp t+\tau, k_x’, y, z\rp}\right] \\
 &= L_{\text{\tiny{tot}}} \int_{\left[-L_{\text{\tiny{tot}}}/2\ ; \ L_{\text{\tiny{tot}}}/2\right]^2} \lp 1-\frac{k_x^2}{k^2}\rp \frac{E_\nu^{\text{\tiny{E}}}(k)}{2} F^{(N)}\lp\frac{\tau}{T_k}\rp \rd k_y \ \rd k_z\\
 &= L_{\text{\tiny{tot}}} \int\limits_0^{+\infty} \frac{\rho^2}{k_x^2+\rho^2}\frac{E^{\text{\tiny{E}}}_\nu\lp\sqrt{k_x^2+\rho^2}\rp}{2} F^{(N)}\lp \frac{\tau}{T_{\sqrt{k_x^2 + \rho^2}}}\rp\ 2\pi\rho\ \rd \rho \ .
\end{aligned}
\end{equation}
This integral expression cannot be written as a function depending on $k_x$ and $\tau$ multiplied by a constant. It can however be used to numerically compute the theoretical prediction $  C_{\text{\tiny{long}}}^{(N),\nu} \lp \tau, k_x\rp $, that are plotted for
reference in Fig. \ref{Fig:TempCorr3D}(b) in solid line.

\section{One-dimensional Fourier modes correlation function and its convergence onto a Gaussian.}
\label{App:1DTempCorr}

\begin{figure}
\centering
\includegraphics[scale = 0.7]{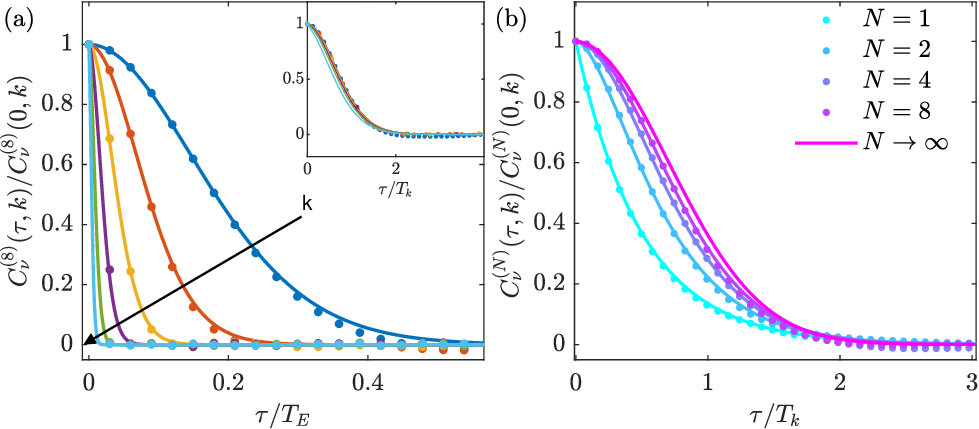}
\caption{One dimensional temporal Fourier mode \cf . (a) 1D Fourier mode \cf\ for the set of Fourier modes $k_n L_{\text{\tiny{tot}}} =7,\ 15,\ 31,\ 63,\ 127,\ 255 $. Solid line curves represent the theoretical predictions $F^{(N)}\lp {\tau}/{T_k}\rp$\eqref{eq:FNTAU} entering in \eqref{eq:1DTempCorr} for the aforementioned Fourier modes while dots are numerical estimations. This simulation has been lead with the same parameters as the 3D one except for the number of layer $N$, here equal to $N=8$. Time is rescaled by $T_k$ in the inset showing that all curves collapse onto a single nearly Gaussian decreasing function. (b) Pointwise convergence of the 1D Fourier mode \cf\ onto a Gaussian for a single Fourier mode $k_n L_\text{\tiny{tot}} = 15$ as a function of $\tau/T_k$ when increasing the number of layer $N$. Solid line are the theoretical predictions $F^{(N)}$ \eqref{eq:FNTAU} for $N = 1,\ 2,\ 4,\ 8$, their pointwise limit $F^{(\infty)}$ \eqref{eq:FInfiniteNTAU} in pink and dots are numerical estimations for the same number of layers $N$.  }
\label{Fig:TempCorr1D}
\end{figure}

We study here the Fourier modes correlation function of a one-dimensional model similar to the previous 3D one \eqref{eq:OU_embedded_N_N} and numerically show their convergence toward a Gaussian. {We define the one-dimensional $N$-layers model exactly the same way as \eqref{eq:OU_embedded_N_N} without a Leray projector and by replacing $E_\nu^{\text{\tiny{E}}}(k)$ by $E_\nu^{\text{\tiny{E,long}}}(k)$}. {Its} one-dimensional Fourier modes correlation function write in the \ssr: 
\begin{equation}
\label{eq:1DTempCorr}
\E\left[\widehat{u}^{(N),\nu}_{1D}(t+\tau,k_n)\overline{\widehat{u}^{(N),\nu}_{1D}(t,k_m)}\right]= L_{\text{\tiny{tot}}}F^{(N)}(\tau/T_{k_n})E_\nu^{\text{\tiny{E,long}}}(k_n)\delta_{n,m},
\end{equation}
with $F^{(N)}$ defined in \eqref{eq:FNTAU}. This result is summarized in Fig. \ref{Fig:TempCorr1D}.

Figure \ref{Fig:TempCorr1D}(a) shows both numerical estimation (dots) and theoretical prediction (solid line) of Fourier modes correlation function for $N = 8$ layers, in the uni-dimensional case
and for log-spaced Fourier modes. The inset shows the exact same curves as a function of $\tau/T_k$ and we observe that all curves collapse on a single nearly Gaussian function. In Figure \ref{Fig:TempCorr1D}(b), we show the convergence of a Fourier mode correlation function toward a Gaussian while increasing the number of layers $N$. We can observe that there is a slight but sufficiently small difference between $N = 8$ and $N\rightarrow +\infty$ to consider a $8$-layers process to be numerically smooth and whose Fourier mode \cf \ are almost Gaussian. 
The model is therefore fully able to depict a Gaussian-like Fourier mode correlation function. To lead those numerical simulations, we have chosen the same set of parameters as for the $3D$ simulations, namely $D_3 = 3.62$, $D_2 = 0.021$, $H = 1/3$, $\beta = 1/2$, $L= L_{\text{\tiny{tot}}} =2\pi$,  $\eta_d =0.085$, $N_x = 1024$, $N_t = 5028$ and $\delta t = 0.002$. The integral time scale $T_E$ has been defined arbitrarily by $T_E = \frac{L}{\sigma}$, where $\sigma^2$ is the variance of the process $\sigma^2 = 2\int\limits_0^{+\infty} E_\nu^{\text{\tiny{E,long}}}(k) \ \rd k$. An ensemble average of the process has been made over 100 realizations to estimate numerically the Fourier modes \cf s \ in Fig. \ref{Fig:TempCorr1D}.

\bibliographystyle{jfm}
%\bibliography{/Users/lchevill/Nextcloud/Redac/MyBibTex/mybiblioJune2015}

\end{document}